\documentclass[useAMS,usenatbib,usegraphicx,fleqn]{mn2e}
\usepackage{threeparttable}
\usepackage{tabularx}
\usepackage{amsmath}

\newcommand{\mnras}{MNRAS}
\newcommand{\aj}{AJ}
\newcommand{\apj}{ApJ}
\newcommand{\apjl}{ApJL}
\newcommand{\apjs}{ApJS}
\newcommand{\nat}{Nature}
\newcommand{\aap}{A\&A}

\newcommand{\aapr}{A\&AR}
\newcommand{\araa}{ARAA}
\newcommand{\pasj}{PASJ}
\newcommand{\pasp}{PASP}

\begin{document}

\title{Shell-Shocked:  The Interstellar Medium Near Cygnus X-1}
\author[Sell et al.]
{\parbox{\textwidth}{P.~H.~Sell$^{1,2}$\thanks{email: paul.sell@ttu.edu},
S.~Heinz$^2$,
E.~Richards$^3$,
T.~J.~Maccarone$^1$,
D.~M.~Russell$^4$,
E.~Gallo$^5$,
R.~Fender$^6$,
S.~Markoff$^7$,
M.~Nowak$^8$}\vspace{0.4cm}\\
$^1$Department of Physics, Texas Tech University, Box 41051, Lubbock, TX 79409, USA \\
$^2$Department of Astronomy, University of Wisconsin-Madison, 475 N. Charter St., Madison, WI, USA \\
$^3$Department of Astronomy, Indiana University-Bloomington, 727 East Third Street, Bloomington, IN 47405, USA \\
$^4$New York University Abu Dhabi, P.O. Box 129188, Abu Dhabi, United Arab Emirates \\
$^5$Department of Astronomy, University of Michigan, 500 Church Street, Ann Arbor, MI 48109, USA \\
$^6$Department of Physics, University of Oxford, Keble Road, OX1 3RH, Oxford, UK \\
$^7$Astronomical Institute ``Anton Pannekoek", University of Amsterdam, PO Box 94249, 1090 GE Amsterdam, The Netherlands \\
$^8$Massachusetts Institute of Technology, Kavli Institute for Astrophysics, Cambridge, MA 02139, USA}

\date{}

\pagerange{\pageref{firstpage}--\pageref{lastpage}} \pubyear{}

\maketitle

\begin{abstract}

We conduct a detailed case-study of the interstellar shell near the high-mass X-ray binary, Cygnus~X-1.  We present new WIYN optical spectroscopic and \emph{Chandra} X-ray observations of this region, which we compare with detailed MAPPINGS III shock models, to investigate the outflow powering the shell.  Our analysis places improved, physically motivated constraints on the nature of the shockwave and the interstellar medium (ISM) it is plowing through.  We find that the shock is traveling at less than a few hundred km~s$^{-1}$ through a low-density ISM ($< 5$~cm$^{-3}$).  We calculate a robust, $3 \sigma$ upper limit to the total, time-averaged power needed to drive the shockwave and inflate the bubble, $< 2 \times 10^{38}$~erg~s$^{-1}$.  We then review possible origins of the shockwave.  We find that a supernova origin to the shockwave is unlikely and that the black hole jet and/or O-star wind can both be central drivers of the shockwave.  We conclude that the source of the Cygnus~X-1 shockwave is far from solved.

\end{abstract}

\label{firstpage}


\section{Introduction}

Cygnus~X-1 is a well-studied, bright high-mass X-ray binary (XRB).  The binary consists of a black hole (BH) in a 5.6-day orbit \citep{bolton72,webster72} accreting mass via a stellar wind from a O9.7ab supergiant star \citep[HD~226868;][]{caballero-nieves09}.  The masses of the BH and O-star are $14.8 \pm 1.0$~M$_{\sun}$ and $19.2 \pm 1.9$~M$_{\sun}$, respectively \citep{orosz11}.

\begin{figure*}
\begin{tabular}{lr}
 \includegraphics[width=0.48\textwidth]{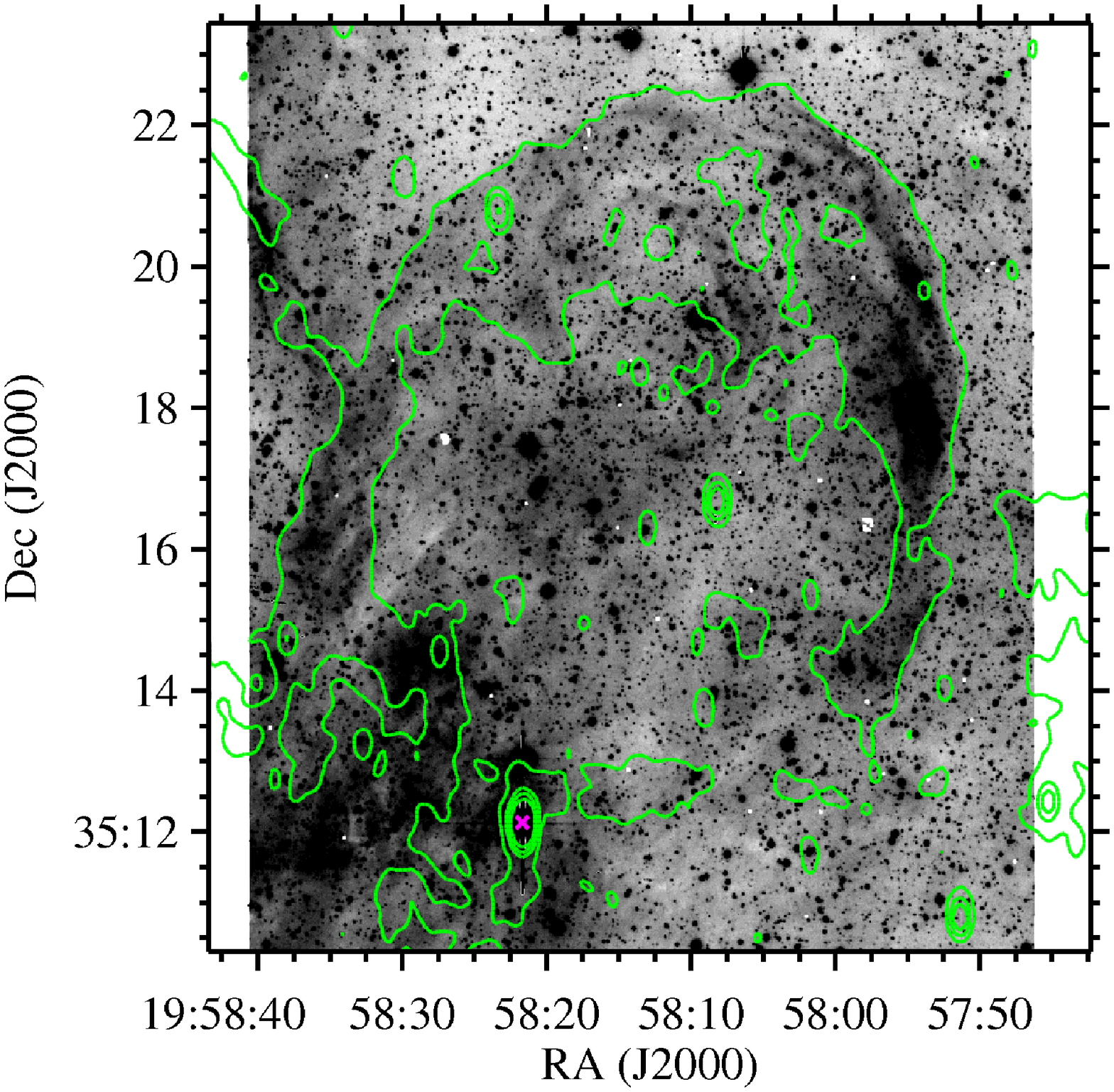}
 &
 \includegraphics[width=0.48\textwidth]{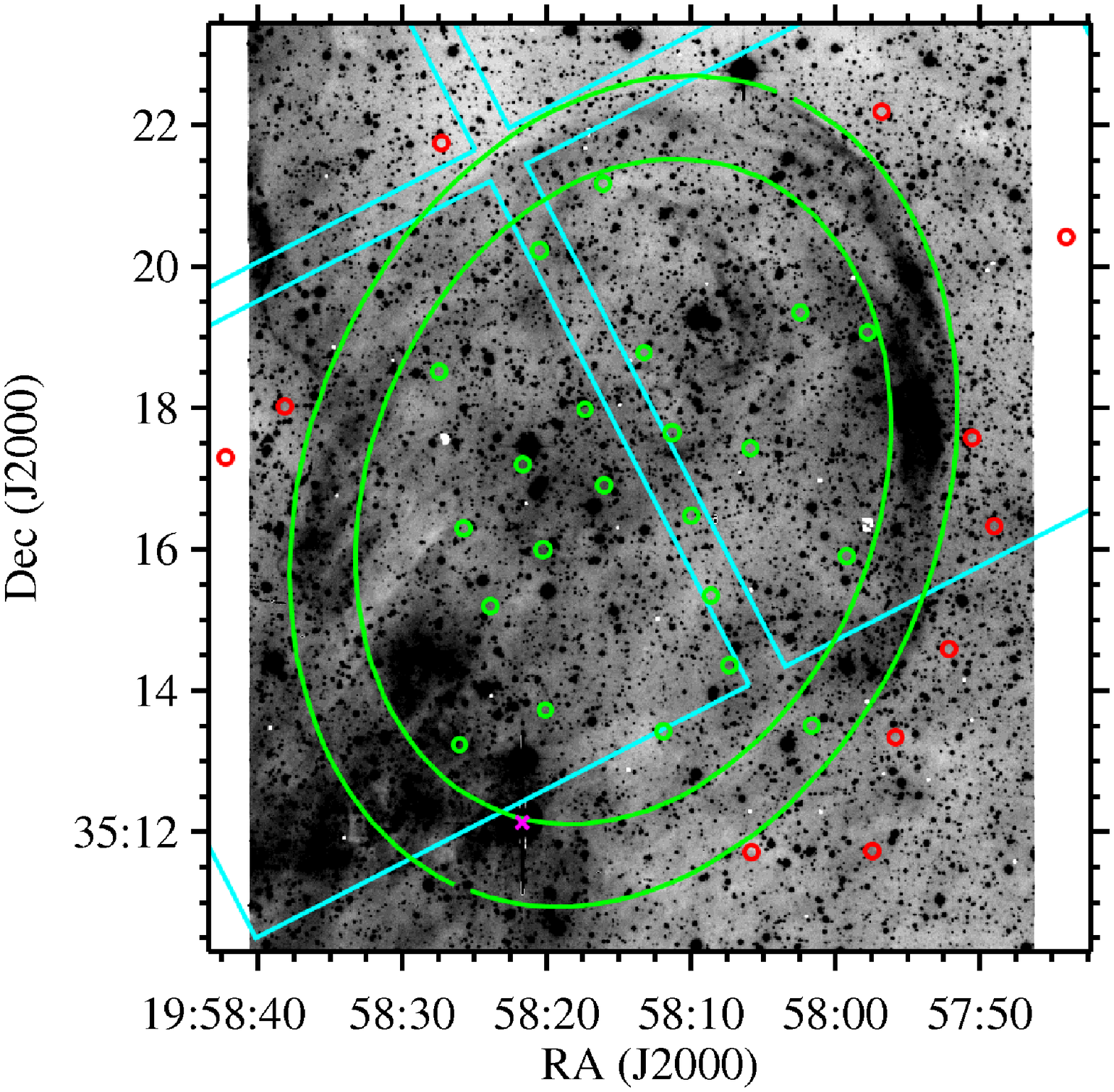}
\end{tabular}
 \caption{H$\alpha$ images from \citet{russell07} with 1.4~GHz radio contours at 0.08, 0.5, 1.0, and 2.0~mJy (left) from \citet[]{gallo05} and other selected regions (right) overlayed.  In both panels, Cygnus~X-1 has been marked by a magenta ``X".  On the right, the small circles show the location and size of our optical fibers, which have been scaled up by 4$\times$ in radius for clarity.  The fibers inside and outside of the limb-brightened shell are colored green and red, respectively.  We also overplot the footprints of the {\em Chandra} ACIS-I CCDs (cyan) and the annular elliptical region (green) that traces the limb-brightened shell.  The limb-brightened elliptical shell and {\em Chandra} ACIS-I CCD footprints are identical in both this figure and Fig.~\ref{fig:Xray_image}.  At the well-determined distance to Cygnus~X-1, 1~arcmin = 0.55~pc.}
 \label{fig:Ha_images}
\end{figure*}

Observations of the diffuse interstellar gas north of Cygnus~X-1 at radio \citep{gallo05} and optical \citep{russell07} wavelengths have revealed a bow shock of what is likely thermally emitting gas expanding into the interstellar medium (ISM) at velocities $\sim 100-360$~km~s$^{-1}$ (see Fig.~\ref{fig:Ha_images}).  As expected in the environments of XRBs \citep{heinz02}, extended nebular optical line emission is seen ([O~III] and H$\alpha$ observed in this case) throughout the projected, limb-brightened bubble that is spatially coincident with the 1.4~GHz radio emission.  While other comparable observations of such emission around XRBs in the Milky Way are limited \citep[e.g.][]{wiersema09}, such emission has been observed on physically larger scales around many ultra-luminous X-ray sources\footnote{Ultra-luminous X-ray sources (ULXs) are a subclass of XRBs with $L_X \ga 10^{39}$~erg~s$^{-1}$, assuming isotropic emission, which is above the Eddington limit for a stellar-mass BH.} \citep[e.g.][]{pakull03}.  However, detailed modeling of the optical line emission in the extended shock and expanding bubbles around XRBs has only been undertaken for a couple of sources:  SAX~J1712.6$-$3739 \citep{yoon11} and S26 in NGC~7793 \citep{dopita12}.

These observations and models are particularly important because they enable estimates of the total power injected by the binary into the ISM, including mechanical work done on the nearby gas.  This is analogous to measuring the total power in supermassive BH jets in galaxy clusters by measuring radio lobes and X-ray cavities \citep[e.g.][]{mcnamara07}.  Measurements of XRB jet power can anchor the kinetic luminosity function of jets down to very low injection energies \citep{merloni08}.  Conducting measurements of stellar-mass BH jets enables us to test the scale-invariance of BH jets \citep{heinz03,merloni03,falcke04} and understand the feedback potential of XRBs \citep{heinz05,fender05,justham12}.  This is particularly relevant for Cygnus X-1, where the accretion flow produces steady, relativistic jets in the low/hard state \citep{pooley99,wilms06}.  Radio observations have resolved the northern compact jet on milliarcsecond scales \citep{stirling01,fender06}, which enables us to constrain jet models more accurately than with other XRBs \citep[e.g.][]{heinz06}.

For these reasons, we have undertaken follow-up observations of the shock front near Cygnus~X-1 with the {\em Chandra X-ray Observatory} and the WIYN 3.5-m telescope.  We combine our observations with previous 1.4~GHz radio measurements \citep{gallo05}, which we will then compare to detailed shock models to place constraints on the shockwave.

In Section~\ref{section:datared}, we describe the reduction of our multiwavelength observations.  In Section~\ref{section:analysis}, we analyze our observations and compare them to detailed MAPPINGS~III shock models.  Finally, we revisit the interpretation of the outflow in Section~\ref{section:discussion} and state our conclusions in Section~\ref{section:conclusions}.  Throughout the paper, we assume a distance to Cygnus~X-1 of 1.9~kpc, which is now known to an accuracy of $\sim 5$ per cent from recent X-ray dust scattering halo and VLBA trigonometric parallax measurements \citep{xiang11,reid11}.

\section{Observations and Data Reduction}
\label{section:datared}

\subsection{X-ray}

In order to search for X-ray emission from the expanding shockwave, we observed the region northwest of Cygnus~X-1 with the {\em Chandra X-ray Observatory} for a continuous 46~ksec on 2007-03-11 (OBS\_ID=7501).  To protect the instrument, Cygnus~X-1 was placed just off {\em Chandra's} ACIS-I2 CCD \citep[see Fig.~\ref{fig:Xray_image};][]{garmire03}.  Data were taken in timed exposure mode and telemetered in faint mode.  Data reduction and analysis were completed using CIAO version 4.3 \citep{fruscione06}, XSpec 12.6.0q \citep{dorman01}, and ACIS Extract version 2011-03-16 \citep[AE;][]{broos10}.  Compared with previous gratings observations, this observation was at least an order of magnitude more sensitive to diffuse flux because of the long exposure and the lack of gratings.

We reprocessed the level=1 event files using the {\sc chandra\_repro} script with ``pix\_adj=none" to apply the most-recent calibration updates available (CALDB version 4.4.3).  We ran {\sc wavdetect} \citep{freeman02} to identify point sources and then excluded them when we used AE to extract diffuse emission in the region where we expected to see the shock based on the H$\alpha$ image from \cite{russell07}.  A monoenergetic exposure map at 1.0~keV was created where needed for {\sc wavdetect} and the AE analysis.  No statistically significant ($>3\sigma$) background flares were detected in this observation.

\subsection{Optical}

\begin{figure*}
 \includegraphics[width=\textwidth]{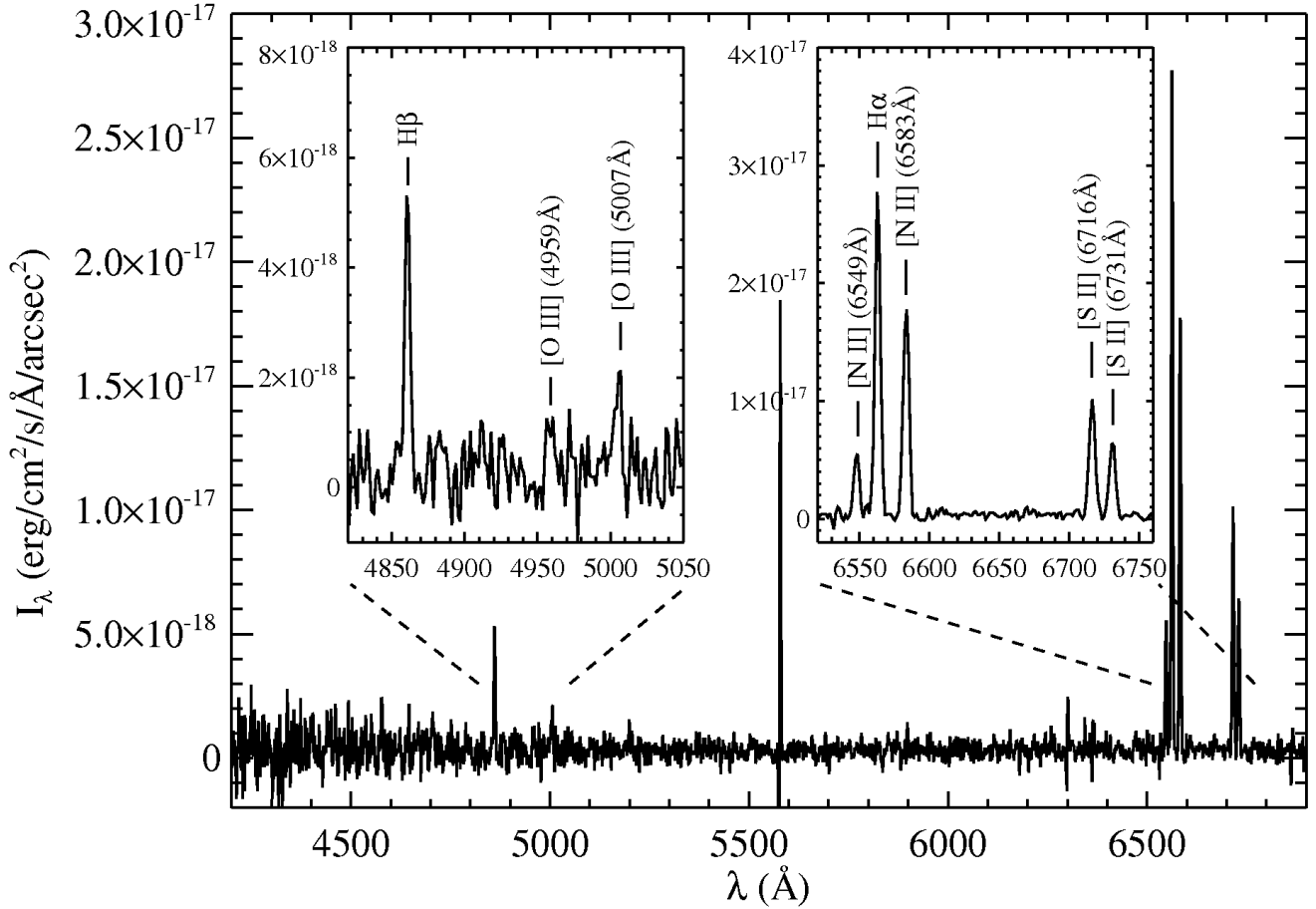}
 \caption{Averaged, background-subtracted spectrum for our WIYN observations.  The two spectral regions that we use in our analysis are magnified, where the relevant emission lines are labeled.  Note that our analysis method uses the individual fiber-by-fiber source and background spectra, not this averaged version.  See text for details.}
 \label{fig:spectra}
\end{figure*}

We observed the same field as the X-ray observations on 2010-07-08 using the Hydra multi-object positioner with the blue-sensitive, $3.1^{\prime \prime}$ fibers \citep{barden95} attached to the upgraded bench spectrograph \citep{bershady08} on the WIYN 3.5-m telescope.  We initially observed the field with over 100 fibers in two different configurations.  The strength of Hydra is that we were able to place fibers in between stars in this considerably crowded field.

We took and applied standard calibrations: biases/zeros, darks, domeflats, skyflats (sensitivity corrections), and arclamps (wavelength calibration).  We used standard IRAF v2.14 \citep{tody93} tools to process the images, including {\sc dohydra} to extract the fibers.  We used two standard star observations with the average extinction at KPNO to calibrate our spectra.  A comparison of the absolute calibration from each of our two standard stars suggests a $\sim 15$ per cent absolute flux uncertainty over the spectral range we analyzed ($\lambda \sim 4800$~--~$6800$~\AA) in addition to measured relative uncertainties.  All quoted flux uncertainties throughout this paper do not include this systematic uncertainty in absolute flux (it does not significantly affect the conclusions from our line ratio analysis, especially since most other uncertainties are much larger).  In addition, our extinction-corrected H$\alpha$ surface brightness shown in Table~\ref{table:line_limits} is consistent with measurements made by \cite{russell07}.

Since we aim to analyze very faint diffuse emission, which requires high data quality, we implemented a strict observing and data reduction strategy.  This strategy led us to reduce the amount of data we include in our analysis significantly.

First, we reject the second configuration from our analysis for two main reasons.  1) Inspection of the data reveals that this configuration yields a factor of two lower signal-to-noise of the nebular emission line flux.  This reduction in signal-to-noise appears to be caused by scattered light from the Moon, which began rising shortly after the start of observations in this configuration.  2) The second configuration also contains a large fraction of fibers far from the center of the shell, which is problematic because we do not correct for limb-brightening in our analysis and because these fibers do not sample the shock along most of its path as the models do (see section~\ref{section:models}).

Second, after visual inspection, we reject approximately half of the fibers in the first configuration for two main reasons.  1) Because the field is so crowded, some of the fibers are contaminated by stray light from a star near the fiber that overwhelms the very faint diffuse line emission used in our analysis.  2) Some of the fibers suffer from the same problem as in the second configuration:  they are far from the center of the shell.

Therefore, our analysis includes 22 fibers inside the limb-brightened shell, which we refer to as source fibers throughout this paper, and 11 fibers outside of the limb-brightned shell, which we refer to as background fibers.  Each of these 33 fibers were observed for 1.3 hours under dark, transparent conditions in one night.  Figure~\ref{fig:Ha_images} shows the positions of these fibers relative to the {\em Chandra} ACIS-I CCDs and projected outermost shock edge, which is highlighted with the same elliptical annular region used in the X-ray analysis in the next section.

Figure~\ref{fig:spectra} shows the averaged, background-subtracted spectrum of these fibers.  The averaged, background-subtracted emission line surface brightness sensitivity is $I = 3$~--~$8 \times 10^{-18}$~erg~cm$^{-2}$~s$^{-1}$~arcsec$^{-2}$ over our wavelength range $\sim 4300$~--~$6900$~\AA ~ ($3\sigma$; less sensitive at bluer wavelengths).  Note that this is not the spectrum from which the emission line fluxes were measured for our final analysis (see section~\ref{section:analysis} for details).

We examined the continua in our observations.  After clipping away lines and extreme values in each case, we found that the mean continuum value within our spectral range (4120\AA~--~6950\AA) is consistent with zero within $1\sigma$.  While it appears that there might be a small upward deviation in Fig.~\ref{fig:spectra}, given the cumulative statistical and systematic (e.g. flatfielding) errors, it is not significant.  Therefore, we cannot make any definitive statements regarding the underlying continuum within the shell.

\section{Analysis}
\label{section:analysis}

\subsection{A Search for Faint, Diffuse X-ray Emission}

\begin{figure}
 \includegraphics[width=0.47\textwidth]{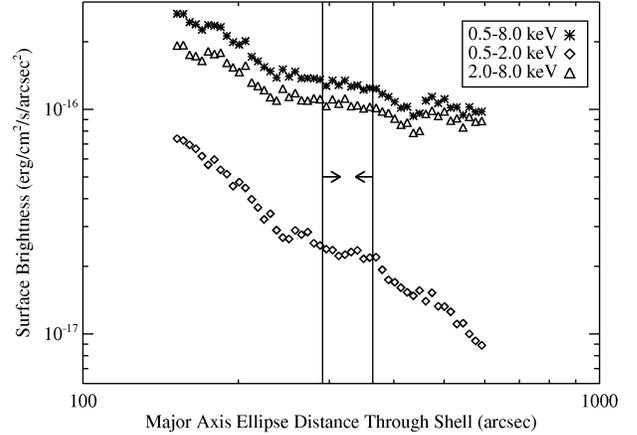}
 \caption{X-ray elliptical profiles through the limb-brightened shell for a few different energy bins.  The errors, estimated from the Poisson error in the counts, are always smaller than the size of the symbols.  The elliptical annuli are centered at the middle of the optical limb-brightened shell at (RA,Dec) = (299.561,35.280) with $b/a = 0.74$ and are rotated clockwise by $21^\circ$, where the major axis, $a$, points north.  Each elliptical annulus samples a range of Cygnus~X-1's extended PSF and varying sections of the four-chip ACIS-I array, which can introduce additional, small systematic uncertainties.  Pixels within $\sim 150^{\prime \prime}$ of Cygnus~X-1 were excluded from the elliptical annuli to help minimize strong contributions from Cygnus~X-1's extended PSF.  The limb-brightened optical shell is at $\sim 290$--$360^{\prime \prime}$.  No X-ray emission from the shock in the shell is present.}
 \label{fig:Xray_elliptical_profile}
\end{figure}

\begin{figure*}
 \includegraphics[width=\textwidth, trim=1in 0.5in 0.5in 1in, clip]{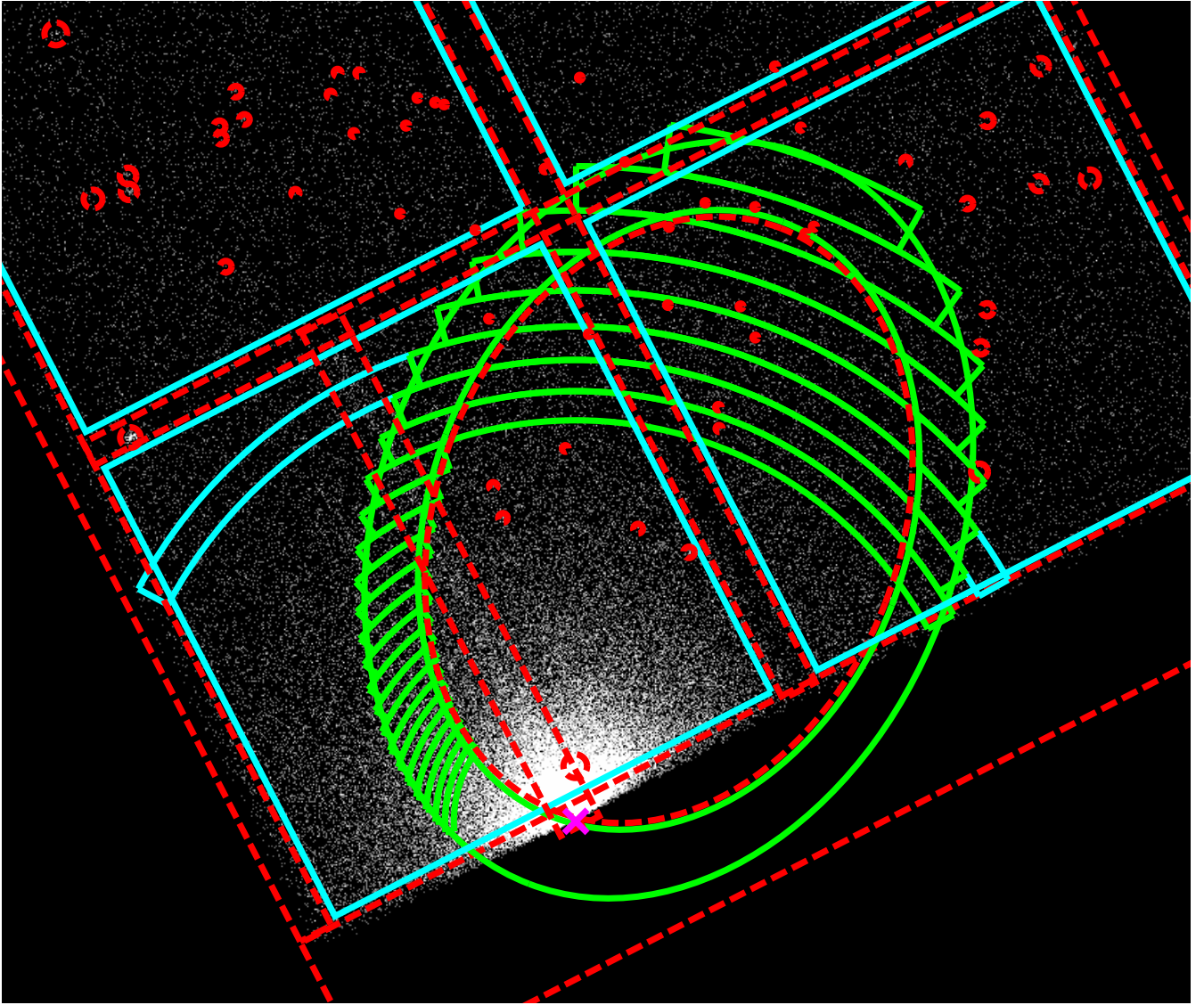}
 \caption{{\em Chandra} unfiltered X-ray image of the level=2 event file with various regions overlayed for reference.  The limb-brightened elliptical shell is highlighted in solid green.  The solid green truncated radial annuli are extraction regions used to measure the radial profile surface brightness.  The solid blue truncated radial annulus is an example background slice corresponding to the green source annulus inside the limb-brightened elliptical shell at the same radius.  The footprints of the ACIS-I CCDs are in cyan.  The limb-brightened elliptical shell and ACIS-I CCD footprints are identical in both this figure and Fig.~\ref{fig:Ha_images}.  The small magenta ``X" denotes the position of Cygnus~X-1, which was placed immediately off the edge of the CCD to prevent damage to the I2 chip.  The red, dashed regions are excluded:  foreground/background point sources, the readout streak from Cygnus~X-1, dithered regions in between and around the chips, and the region interior to the limb-brightened shell.  North is up and east is left.}
 \label{fig:Xray_image}
\end{figure*}

\begin{figure}
 \includegraphics[width=0.47\textwidth]{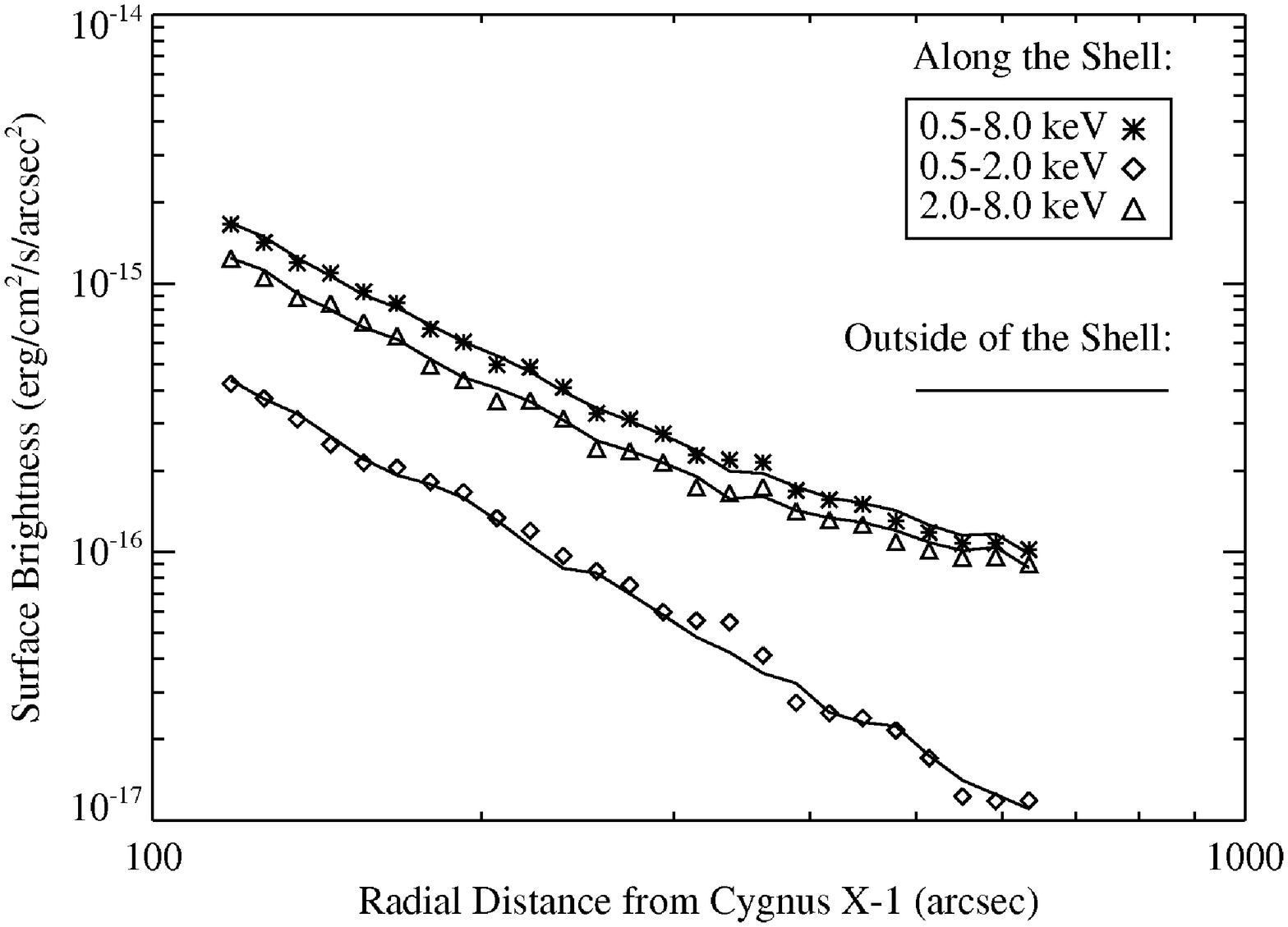}
 \caption{X-ray radial profiles (centered at Cygnus~X-1) for a few different energy bins both on and outside of the limb-brightened shell.  The errors, estimated from the Poisson error in the counts, are always smaller than the size of the symbols.  Each point corresponds to a truncated green annulus inside the limb-brightened elliptical shell in Fig.~\ref{fig:Xray_image}.  The solid lines at each point correspond to the truncated blue annuli outside of the limb-brightened elliptical shell (background regions) in Fig.~\ref{fig:Xray_image}.  The annuli sample varying sections of the four-chip ACIS-I array, introducing additional, small systematic uncertainties.  No X-ray emission from the shock in the shell is present.}
 \label{fig:Xray_radial_profile}
\end{figure}

We searched for diffuse X-ray emission from the outflow throughout the image by making fluxed images (using the CIAO task {\sc fluximage}) and heavily smoothed images, both at multiple energy bands.  We also extracted both radial and elliptical regions across the ACIS-I CCDs.  In all parts of our diffuse source analysis, we construct weighted ARFs with {\sc mkwarf} that take into account variations in the effective area and detector quantum efficiency to a high level of accuracy.  Remaining in the profiles are small ($\la$ a few per cent) systematic errors that we do not attempt to take into account.  Our analysis of our ARFs and inspection of the surface brightness profiles described in the next two paragraphs suggests that this error is relatively small and does not significantly affect our ability to detect the limb-brightened shell.

We created an elliptical profile from a set of logarithmically-spaced elliptical annuli from the center of the limb-brightened shell (Fig.~\ref{fig:Xray_elliptical_profile}).  The dimensions of the semi-major axes of the inner and outer edges of the elliptical shell determined by eye are $3.61^{\prime} \times 4.86^{\prime}$ and $4.50^{\prime} \times 6.07^{\prime}$, respectively, which are rotated $21^{\circ}$ clockwise from north.  We used all of the masks shown in Fig.~\ref{fig:Xray_image} to exclude the dithered regions in between and around the chips, the readout streak, and the background/foreground point sources.  For this profile, we also excluded regions within $\sim 150^{\prime \prime}$ of Cygnus~X-1, where the contamination of the PSF and dust-scattering halo are greatest.

We created a radial profile from a set of logarithmically-spaced radial annuli from Cygnus~X-1, which extend outward along or outside of the limb-brightened shell (Fig.~\ref{fig:Xray_radial_profile}).  Because the limb-brightened shell is projected elliptically on the sky, because the CIAO tasks only accept certain region types, and because regions can only be easily excluded from the inside-out, we extracted diffuse regions in the shapes of truncated, logarithmically-spaced, radial annuli along and outside of the shell.  These annuli and the masked regions are shown in Fig.~\ref{fig:Xray_image}.

In both of these profiles, {\em we do not detect any diffuse emission}.  Based on this non-detection, we derive a $3\sigma$ upper limit to the 0.5--8.0~keV X-ray surface brightness.  To do this, we generated XSpec table models from the MAPPINGS~III models for the full range of parameters we explored (see Table~\ref{table:model_params}).  After jointly fitting the data to these models included with the standard photoabsorption model, {\sc phabs} ($N_H = 5 \times 10^{21}$~cm$^{-2}$), we find $I_{0.5-8.0~keV} \la 1 \times 10^{-18}$~erg~cm$^{-2}$~s$^{-1}$~arcsec$^{-2}$ when taking the most conservative upper limit.  Assuming this upper limit for the surface brightness holds across the detectable part of the limb-brightened shell, we calculate the corresponding flux and luminosity upper limits summed over the shell:  $F_{0.5-8.0~keV} \la 1 \times 10^{-13}$~erg~cm$^{-2}$~s$^{-1}$ and $L_{0.5-8.0~keV} \la 5 \times 10^{31}$~erg~s$^{-1}$.

\subsection{Optical Spectroscopic Analysis}
\label{section:optical_analysis}

\begin{figure}
 \includegraphics[trim=0.0in 0.0in 0.0in 0.1in,clip,width=0.47\textwidth]{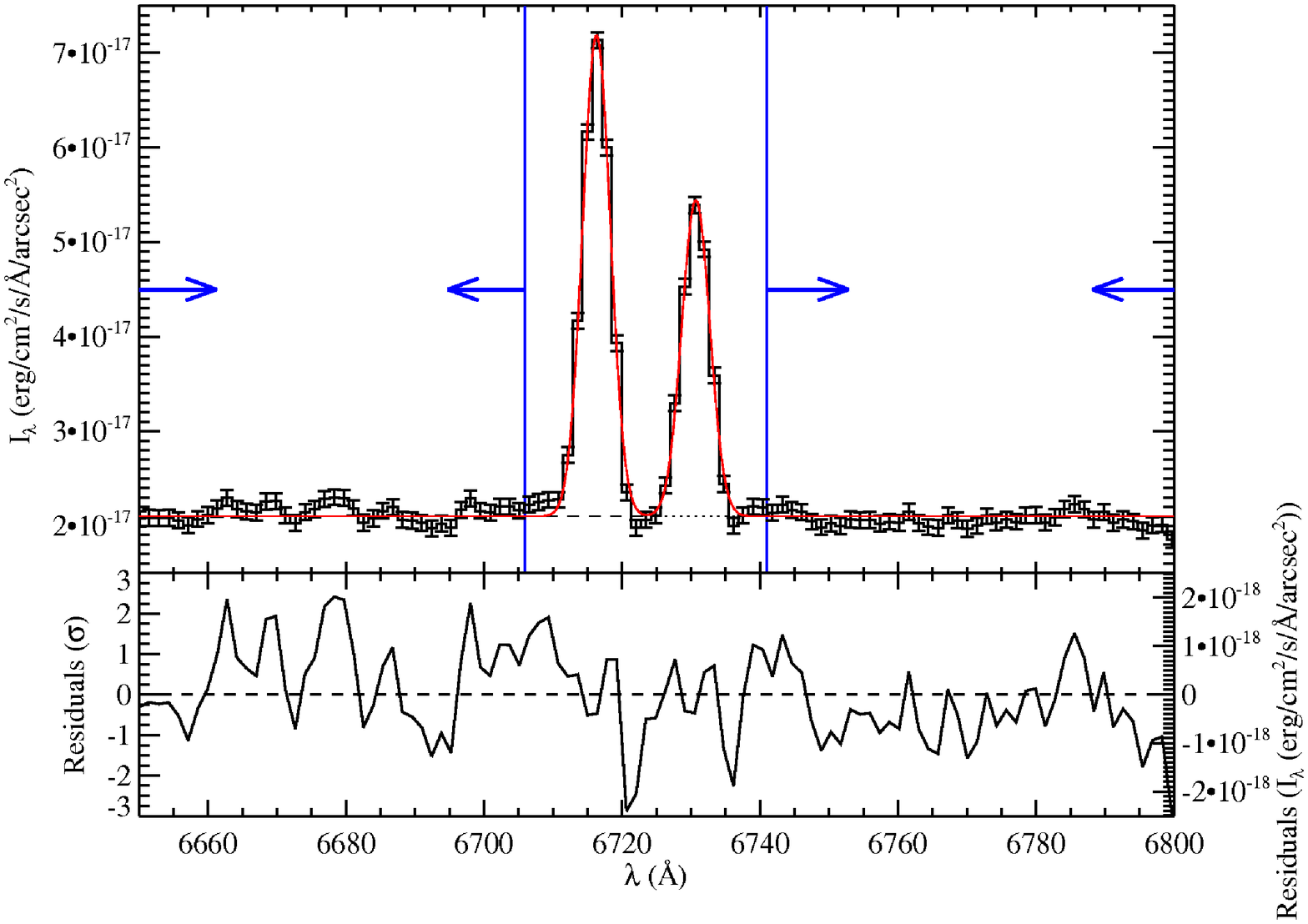}
 \caption{An example fit (red line) to the [S~II] lines in one of the source fibers.  The uncertainty is estimated from the continuum on either side of the lines, which is bounded by blue vertical lines and arrows.  To derive confidence intervals from our Monte Carlo simulations, fits were completed on the source and background spectra separately for each fiber.}
 \label{fig:fitting}
\end{figure}

\begin{figure*}
\begin{tabular}{lr}
\begin{minipage}{0.48\textwidth}
 \includegraphics[trim = 1.25in 0.0in 0.70in 0.0in, clip, width=\textwidth]{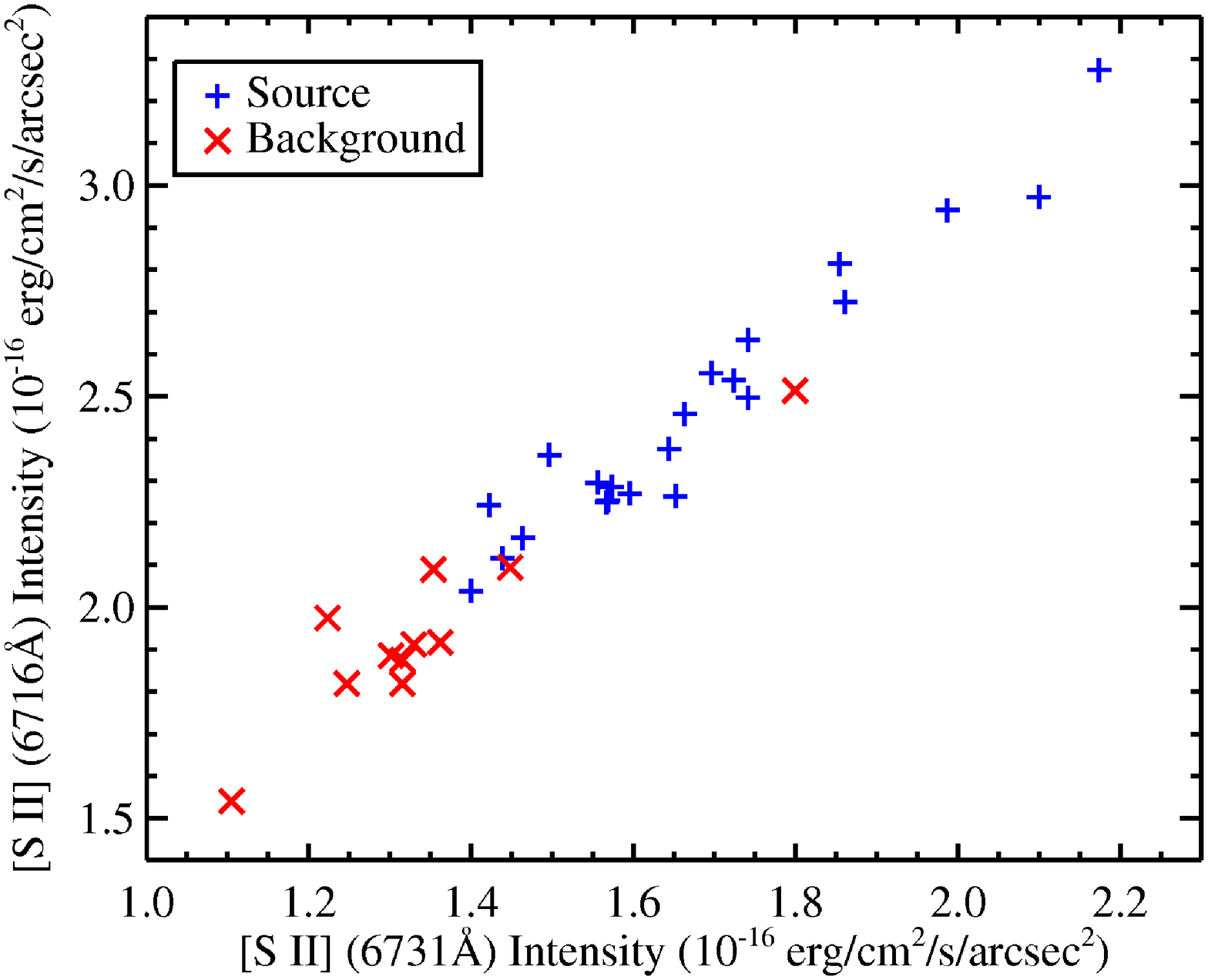}
\end{minipage}
&
\begin{minipage}{0.48\textwidth}
 \includegraphics[trim = 1.25in 0.0in 0.70in 0.0in, clip, width=\textwidth]{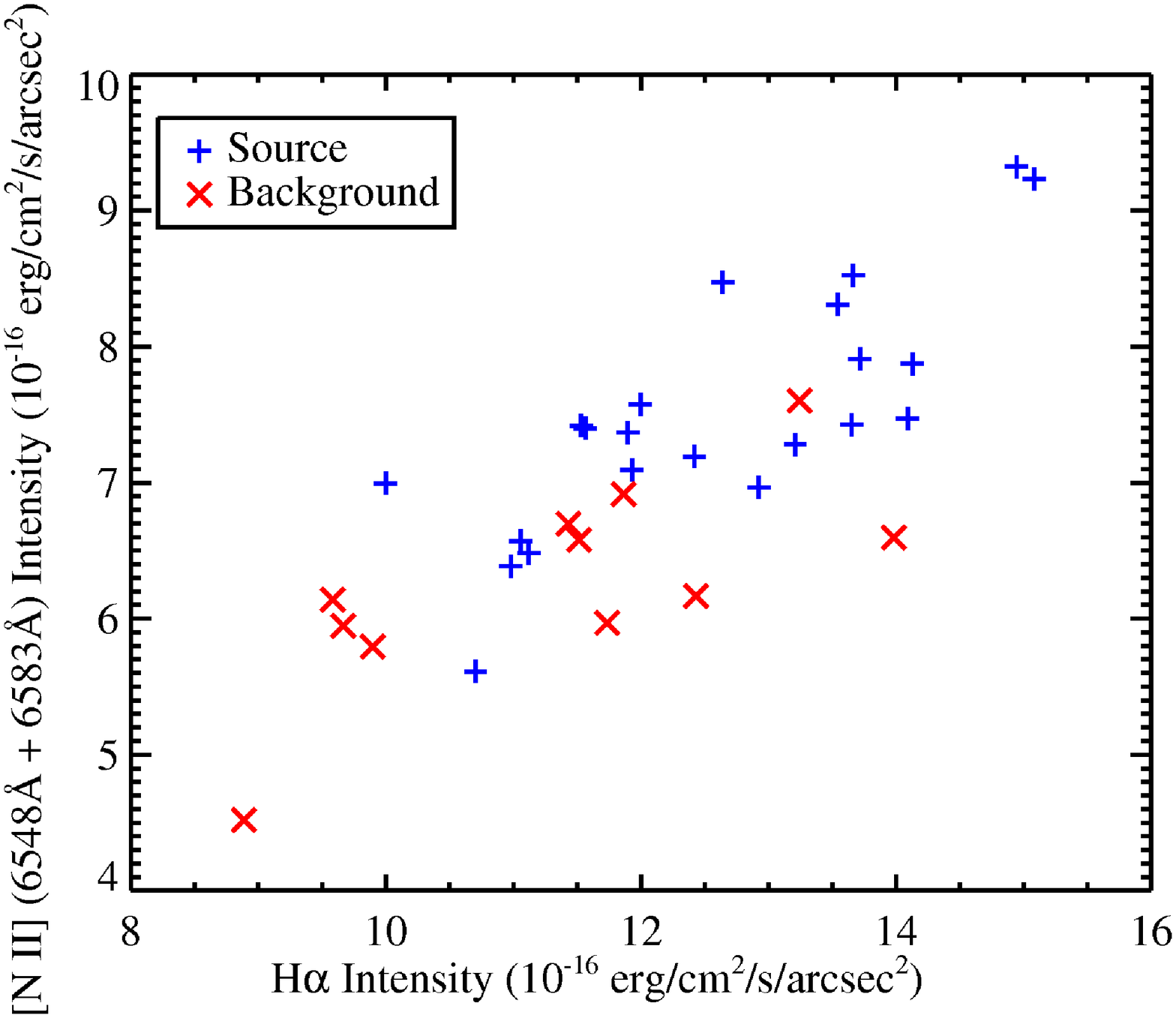}
\end{minipage}
\\
\begin{minipage}{0.48\textwidth}
 \includegraphics[trim = 1.25in 0.0in 0.70in 0.0in, clip, width=\textwidth]{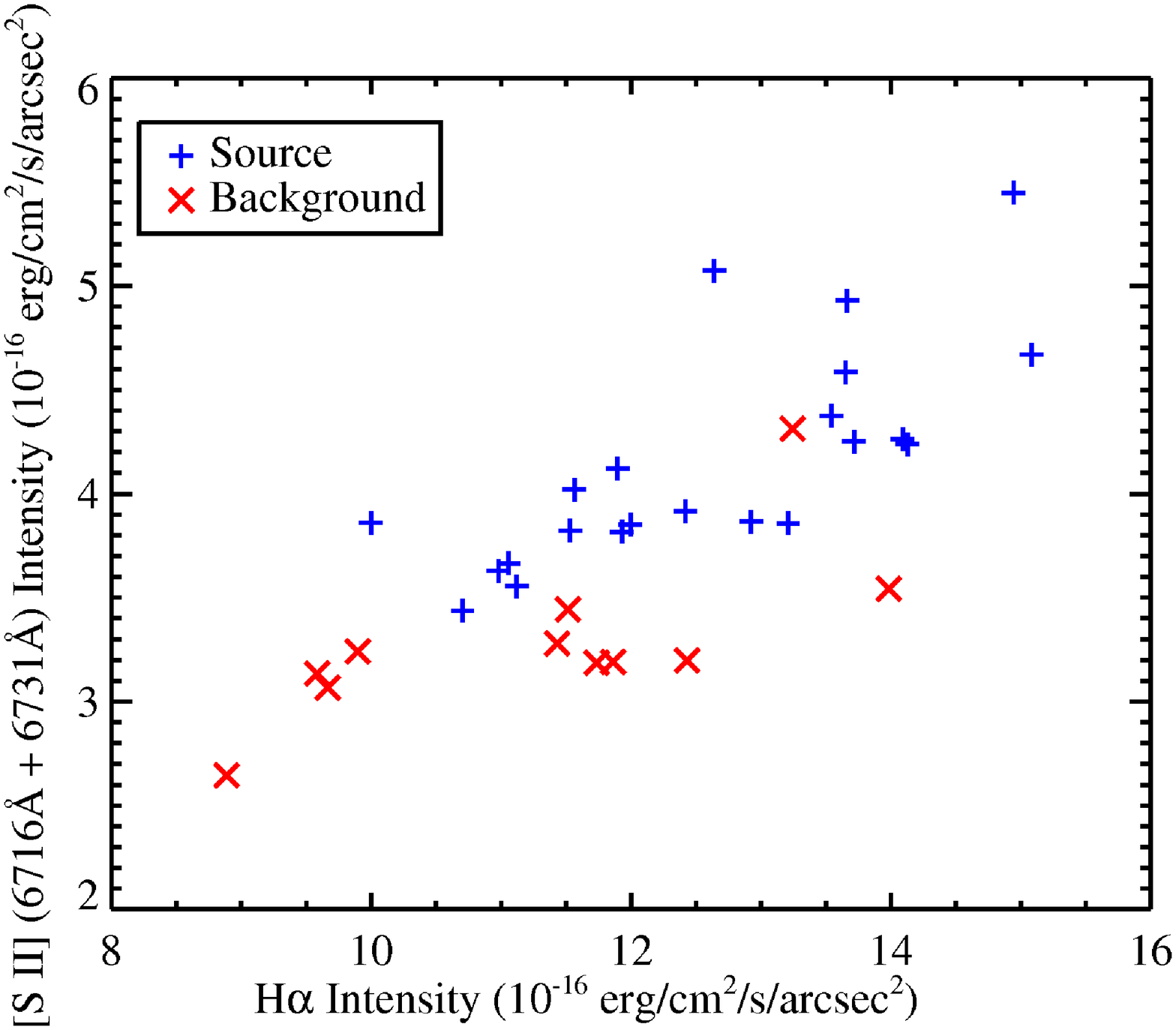}
\end{minipage}
&
\begin{minipage}{0.48\textwidth}
 \includegraphics[trim = 1.25in 0.0in 0.70in 0.0in, clip, width=\textwidth]{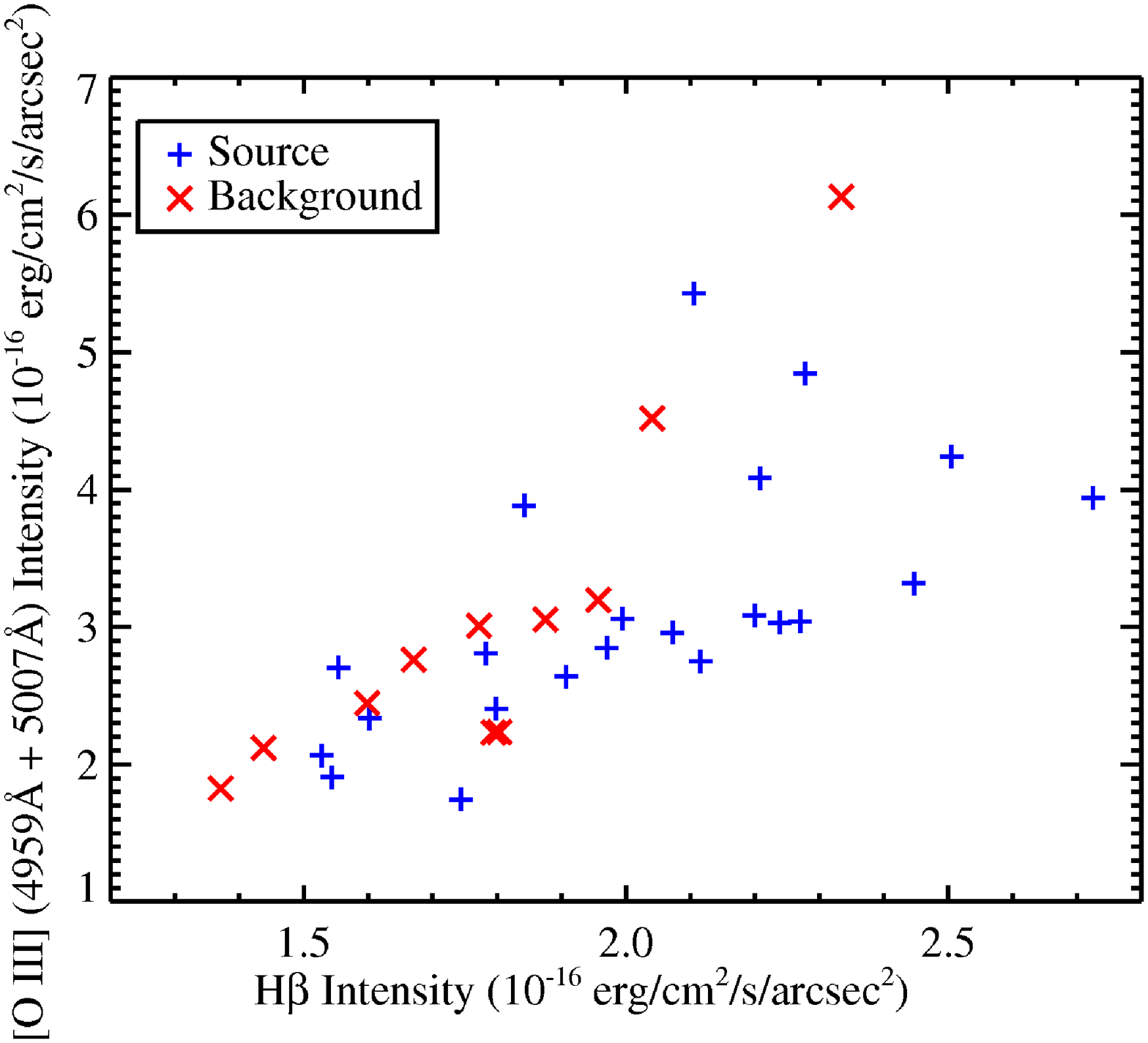}
\end{minipage}
\end{tabular}
 \caption{Distributions of the emission line measurements for each fiber, plotted for each line ratio diagnostic.}
 \label{fig:measurements}
\end{figure*}

We compare measured and shock model emission-line and continuum diagnostics (see section \ref{section:models}) to constrain shock parameters.  Because this requires careful measurements of the emission line intensities and continuum, we jointly fit the emission lines and continuum in each of three spectral ranges containing the H$\beta$ and doublet [O~III] lines (4800~--~5050\AA), H$\alpha$ and the doublet [N~II] lines (6480~--~6650\AA), and doublet [S~II] lines (6650~--~6800\AA).  We fit each line in each of the source and background fibers with a single Gaussian, which provided a good measure of the line fluxes.  An example fit to the [S~II] lines in one of the source fibers is shown in Fig.~\ref{fig:fitting}.  The measured intensities used for our line diagnostics in each source and background fiber (not extinction-corrected) are provided in Fig.~\ref{fig:measurements}.  Our average observed (not extinction-corrected) intensities and uncertainties for the emission lines in the spectrum are provided in Table~\ref{table:line_SBs}.

\begin{table*}
  \centering
  \caption{Average Observed Optical Line Intensities}
  \label{table:line_SBs}
  \begin{threeparttable}
    \begin{tabular}{llll} \hline \hline
      Line                     & Source                   & Background               & Source $-$ Background \\
                               & \multicolumn{2}{c}{($10^{-16}$~erg/cm$^2$/s/arcsec$^2$)} & ($10^{-18}$~erg/cm$^2$/s/arcsec$^2$) \\ \hline
      H$\beta$~(4861~\AA)      & $2.02 \pm 0.09 \pm 0.07$ & $1.79 \pm 0.12 \pm 0.08$ & $22.6 \pm 1.6$ \\
      {[}O~III]~(4959~\AA)$^a$ & $0.75 \pm 0.11 \pm 0.05$ & $0.71 \pm 0.12 \pm 0.10$ & $6.1  \pm 4.4$ \\
      {[}O~III]~(5007~\AA)$^a$ & $2.39 \pm 0.12 \pm 0.15$ & $2.34 \pm 0.09 \pm 0.28$ & $9.4  \pm 2.0$ \\
      {[}N~II]~(6548~\AA)      & $2.70 \pm 0.41 \pm 0.06$ & $2.37 \pm 0.41 \pm 0.07$ & $26.2 \pm 1.6$ \\
      H$\alpha$~(6563~\AA)     & $12.6 \pm 0.36 \pm 0.30$ & $11.3 \pm 0.31 \pm 0.49$ & $128.6 \pm 1.3$ \\
      {[}N~II]~(6583~\AA)      & $4.80 \pm 0.47 \pm 0.14$ & $3.89 \pm 0.47 \pm 0.17$ & $92.2 \pm 1.4$ \\
      {[}S~II]~(6716~\AA)      & $2.47 \pm 0.05 \pm 0.07$ & $1.95 \pm 0.06 \pm 0.07$ & $51.8 \pm 1.6$ \\
      {[}S~II]~(6731~\AA)      & $1.68 \pm 0.05 \pm 0.04$ & $1.35 \pm 0.06 \pm 0.05$ & $32.7 \pm 1.0$ \\ \hline
    \end{tabular}
      \begin{tablenotes}
		\item Column 1.  The emission lines of interest of our spectral range.  Columns 2 and 3.  Averaged (over our fibers) measured surface brightnesses inside and outside of the the limb-brightened shell, source and background, respectively.  The first uncertainty is the statistical error from Gaussian fits to the spectra.  The second uncertainty is the standard deviation of the mean of the surface brightnesses of the lines in the spectra.  Column 4. The surface brightnesses in the background-subtracted spectrum (Fig.~\ref{fig:spectra}).  The errors in column 4 are only the statistical errors from the Gaussian fit to the background-subtracted spectrum, which strongly underestimate the total error.  The total error also includes differences in the fluxes from fiber to fiber and the extinction correction.  When we carefully account for these uncertainties using bootstrap Monte Carlo simulations, we find a {\em much} larger error in the final, extinction-corrected flux (e.g. compare the relative H$\alpha$ uncertainty above to that in Table~\ref{table:line_limits}).  No surface brightnesses in this table have been corrected for galactic foreground extinction.  
		\item [a] The measured strength of a line near the sensitivity limit ($\la$ a few $\sigma$) is likely to be overestimated \citep{rola94}.
      \end{tablenotes}
  \end{threeparttable}
\end{table*}

As is evident in Table~\ref{table:line_SBs}, the background-subtracted line emission from the post-shock region is only a small fraction (4~--~20 per cent) of the emission in the background fibers.  Given how faint some of the background-subtracted lines are, we calculate the probability that the distributions of the source and background surface brightnesses are drawn from the same parent distribution (the null hypothesis, which would suggest that the measured lines are only statistical fluctations).  We reject the null hypothesis if $p < 0.05$.  For the [S~II] ($\lambda \lambda 6716, 6731$~\AA) lines, $p = 3.5 \times 10^{-5}$, $8.6 \times 10^{-5}$.  For the [N~II] ($\lambda \lambda 6548, 6583$~\AA) lines, $p = 2.6 \times 10^{-3}, 4.0 \times 10^{-4}$.  For H$\alpha$ and H$\beta$, $p = 0.025,0.052$.  Finally, for [O~III] ($\lambda \lambda 4959, 5007$~\AA), $p = 0.68,0.86$.  This suggests that all of our line surface brightness measurements are significantly and consistently measureable above the background except for the [O~III] and H$\beta$ lines.  This is consistent with the large uncertainties in H$\beta$ and [O~III] shown in Table~\ref{table:line_SBs}.

Because the observations are dominated by other nebulosity not associated with the shockwave, the uncertainties listed in Table~\ref{table:line_SBs} are relatively large.  The uncertainty in the extinction through the Galactic disk to the shock \citep[$A_V = 2.95 \pm 0.53$~mag, the values adopted throughout this paper;][]{predehl95,russell07} is also significant.  This is further complicated because the uncertainties are correlated when taking common line ratios (see Fig.~\ref{fig:measurements}).  Estimating accurate confidence intervals is essential to constraining the model predictions.

Therefore, instead of estimating the uncertainties using quadrature sums, we used bootstrap Monte Carlo simulations to calculate the errors in the line ratios and extinction-corrected line surface brightnesses for the necessary line diagnostics.  To replicate the distribution of our observed values, we generated $10^7$ random background-subtracted intensity ratios or (extinction-corrected) intensities from Gaussian distributions.  At each iteration, we randomly sampled the 22 source and 11 background fibers independently.  Then we simulated additional noise consistent with the statistical uncertainties from the Gaussian fits to the emission lines (Table~\ref{table:line_SBs}).  Finally, we calculated the background-subtracted intensity ratio or (extinction-corrected) intensity.  In the case of the H$\alpha$ background-subtracted intensity, we calculated the extinction-corrected background-subtracted intensity by drawing random deviates from a Gaussian distribution for $A_V = 2.95 \pm 0.53$~mag assuming $R_V = 3.1$ \citep{cardelli89,predehl95,russell07}\footnote{We considered using the Balmer decrement to constrain the extinction to the nebula but found that the combined uncertainties in the observed and model ratios together were too large to place a better constraint on the extinction.}.

The confidence intervals for the line ratios and surface brightnesses used in our comparisons to shock models are provided in Table~\ref{table:line_limits}.

\begin{table*}
  \centering
  \caption{Optical Line Ratio/Surface Brightness Confidence Intervals From Bootstrap Monte Carlo Simulations}
  \begin{threeparttable}
    \begin{tabular}{lllll} \hline \hline
      Line Diagnostic                                           & Peak Ratio/Intensity  & $1\sigma$ limits & $2\sigma$ limits & $3\sigma$ limits  \\ \hline
      {[}O~III / H$\beta$]~((4959~\AA~+~5007~\AA)~/~4861~\AA)   & 1.60                  & -1.93, 1.80       & -13.20, 4.50      & -174.67, 180.35    \\
      {[}N~II / H$\alpha$]~((6548~\AA~+~6583~\AA)~/~6563~\AA)   & 0.87                  & 0.65, 1.48        & 0.36, 3.57        & -27.50, 32.92      \\
      {[}S~II / H$\alpha$]~((6716~\AA~+~6731~\AA)~/~6563~\AA)   & 0.59                  & 0.49, 1.02        & 0.39, 2.58        & -21.76, 25.12      \\
      H$\alpha$~(6563~\AA; $10^{-15}$~erg/cm$^2$/s/arcsec$^2$)  & 0.86                  & 0.51, 1.91        & 0.11, 3.21        & -0.39, 5.27        \\
      {[}S~II]~(6716~\AA~/~6731~\AA)                            & 1.54                  & 1.41, 1.74        & 1.26, 2.03        & 1.12, 2.63         \\ \hline
    \end{tabular}
      \begin{tablenotes}
		\item [a] H$\alpha$ has been corrected here for galactic foreground extinction using $A_V = 2.95 \pm 0.53$~mag \citep{cardelli89,predehl95,russell07}.
      \end{tablenotes}
  \end{threeparttable}
  \label{table:line_limits}
\end{table*}

\subsection{Comparisons to MAPPINGS III Shock Models}
\label{section:models}

Our combined set of observational constraints consists of the existing
radio flux measurement of the shell \citep{gallo05}, the nebular
line diagnostics derived from our WIYN observations, and the
X-ray upper limit.

In order to derive physically meaningful constraints from the data, it
is critical to use accurate shock models because, in the temperature
and density range expected for the shock, the cooling time can be
orders of magnitude shorter than the dynamical time, while the
ion-electron equilibration time can be considerable.  Simple textbook
equilibrium estimates of density and temperature diagnostics derived
from the line ratios are not applicable.

To calculate accurate line and continuum fluxes, we adapted the MAPPINGS~III
shock and photo-ionization code \citep{allen08} to
construct an extensive set of shock models.  MAPPINGS~III
includes non-equilibrium microphysics and fully resolves radiative
cooling behind the shock.  It includes pre-ionization from the
radiative precursor and an arbitrary external source of ionizing
photons in the plane-parallel approximation.

We performed calculations using the adaptive timestep implementation,
referred to in the code as the S4 model.  This model advances
the shock forward in time increments $\Delta t$, chosen such that
$\Delta t \leq C\,t_{\rm min}$, where $C \la 0.1$ is an
adjustable constant and $t_{\rm min}$ is the minimum of the cooling
time, collision time, photo-ionization time, and the recombination
time in any of the shells accumulated behind the shock.

In order to further increase the fidelity of emission behind the
shock, we implemented a time step limiter adopted from hydro-dynamics
methods, which restricts increases in the time step to a fixed ratio
$\Delta t_{i+1} = r \Delta t_{i}$, with $r$ chosen to be between 1 and
2, and with a small initial guess for the starting time step.  This
eliminates the possibility that the first time step under-resolves the
ionization time behind the shock, which occurred in a sub-set of the
initial MAPPINGS~III runs.

Each MAPPINGS~III run integrates the shock\footnote{The shock is integrated
along its entire path length; this is one of the two main reasons why fibers
near the edge of the shell were not included in our analysis (the other is that
we do not correct for limb-brightening).} out to a final
distance, usually taking between $\la100$ (in cases where the gas is in
equilibrium where the microphysical time scales are long compared to
the dynamical time) to $\la1000$ time steps.  We modified
the code to generate integrated spectra and line fluxes for each
time/distance step. This is necessary because the actual distance from
the center of the bubble to the shell along the line of sight depends
on the inclination and is thus poorly constrained.  This approach
allows us to explore a range of shock travel distances, properly
resolved by the code.

Because model runs are slow, iteratively fitting MAPPINGS~III
models to the data was not practical.  Instead, we constructed a grid
of models that span all relevant parameters.  We explored ISM densities
in the range of $0.1\,{\rm cm^{-3}} \leq n_{\rm ISM} \leq 100\,{\rm
  cm^{-3}}$ \citep{cox05} and shock velocities in the range of $40\,{\rm km\,s^{-1}}
\leq v \leq 400\,{\rm km\,s^{-1}}$, both sampled on a 32 by 32
grid.  We further considered pre-shock magnetic fields of $0.1, 0.33,
1.0, 3.33$, and $10\,{\rm \mu G}$ \citep{beck96}, metal abundances
of $0.4$, $1.0$, and $2.0$ times the solar values, and a range of ionizing fluxes from
the O-star companion, reflecting the lack of knowledge of the mean
distance of the shell from the O-star, due to the poorly constrained inclination.
We considered a range of distances from Cygnus~X-1 to the shock of 2~--~11~pc
(up to twice the projected maximum distance of the shell), allowing for a factor
of two in foreshortening due to inclination effects.

This resulted in a total set of 76800 separate model runs, each
spanning a range of distances.  We performed these calculations on
three nodes of the {\em Hydra} cluster in the UW-Madison Astronomy
Department, each capable of executing 16 runs in parallel.  The
aggregate computing time was approximately 6000 CPU hours.

We are interested primarily in the constraints on the ISM density and
shock velocity, because these are the parameters needed to determine
the average power of the outflow that drives the expansion of the
shell.  The ISM density and shock velocity are thus direct probes of the
fluid mechanical properties of the outflow.  We marginalized over the
other parameters (they could not be constrained with the current data).
We show results from our analysis in the
$v_{\rm shock}$--$n_{\rm ISM}$ parameter plane.

Given the detailed likelihood functions of the nebular diagnostics
derived from our Monte Carlo treatment,
we proceeded to derive a quantitative estimate of how well the model
can describe the set of data at our disposal.  In order to derive the
probability density distribution for the X-ray flux, we converted the
output X-ray spectrum from each MAPPINGS~III model grid point
into an XSPEC table model.  We then converted the resulting
$\Delta\chi^2$ into a probability using the $\chi^2$ distribution
function.  To derive the probability
density distribution from the radio luminosity, we integrated
the gaussian distribution using the mean and standard deviation from \cite{gallo05}.

\begin{table}
  \centering
  \caption{MAPPINGS III Model Parameters}
  \label{table:model_params}
  \begin{threeparttable}
    \begin{tabular}{lll} \hline \hline
    	Parameter                  & Parameter Ranges/Values                     \\ \hline
    	Pre-shock velocity (v)$^a$ & 40~--~400 km~s$^{-1}$ in 32 bins$^b$        \\
    	Pre-shock density (n)      & 0.1~--~100 cm$^{-3}$ in 32 bins$^b$         \\
    	Magnetic field (B)         & 0.1, 0.333, 1.0, 3.33, 10.0 $\mu$G          \\
    	Abundance (Z)              & 0.4, 1.0, 2.0 M$_\odot$                     \\
    	O-star Ionizing Flux$^c$   & $1/16$, $1/8$, $1/4$, $1/2$, $1$            \\
    	Distance$^d$               & 2~--~11 pc                               \\ \hline
 	\hline
    \end{tabular}
      \begin{tablenotes}
		\item [a] This is the shock velocity measured in the lab frame.
		\item [b] These bins are logarithmically-spaced.
		\item [c] The fraction of the ionizing O-star flux at a distance of 5.5~pc from Cygnus~X-1.
		\item [d] The distance from Cygnus X-1 to the shock.  The minimum distance that any part of the shock is observed if we assume that the outflow is not inclined to the line of sight is $\approx 2$~pc.  We considered up to 11~pc, $2\times$ the maximum distance to the shock to compensate for inclination effects.
      \end{tablenotes}
  \end{threeparttable}
\end{table}

\begin{figure*}
\begin{tabular}{lcr}
\begin{minipage}{0.31\textwidth}
 	\includegraphics[trim = 0.8in 0.0in 0.4in 0.0in, clip, width=\textwidth]{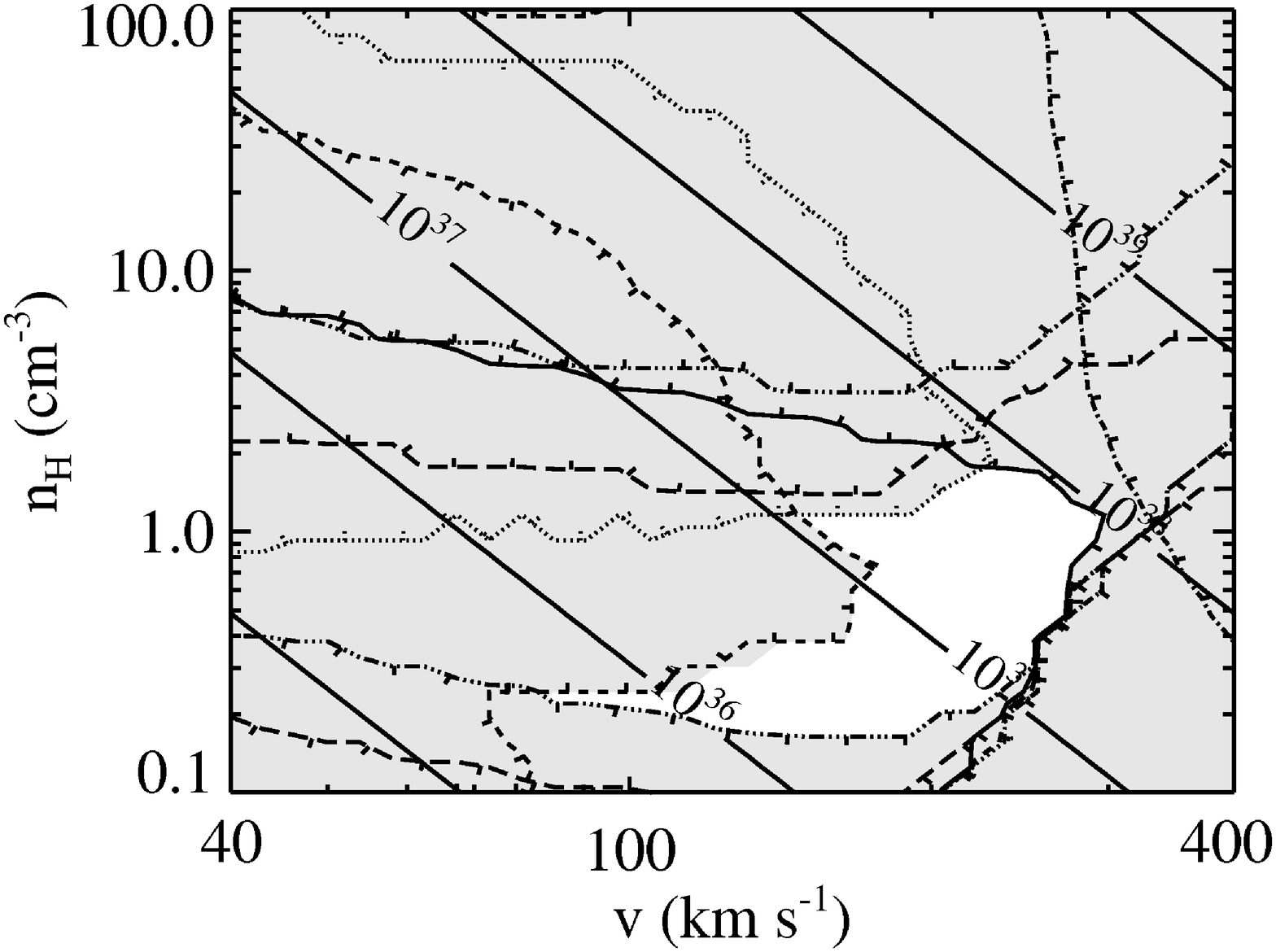}
\end{minipage}
&
\begin{minipage}{0.31\textwidth}
 	\includegraphics[trim = 0.8in 0.0in 0.4in 0.0in, clip, width=\textwidth]{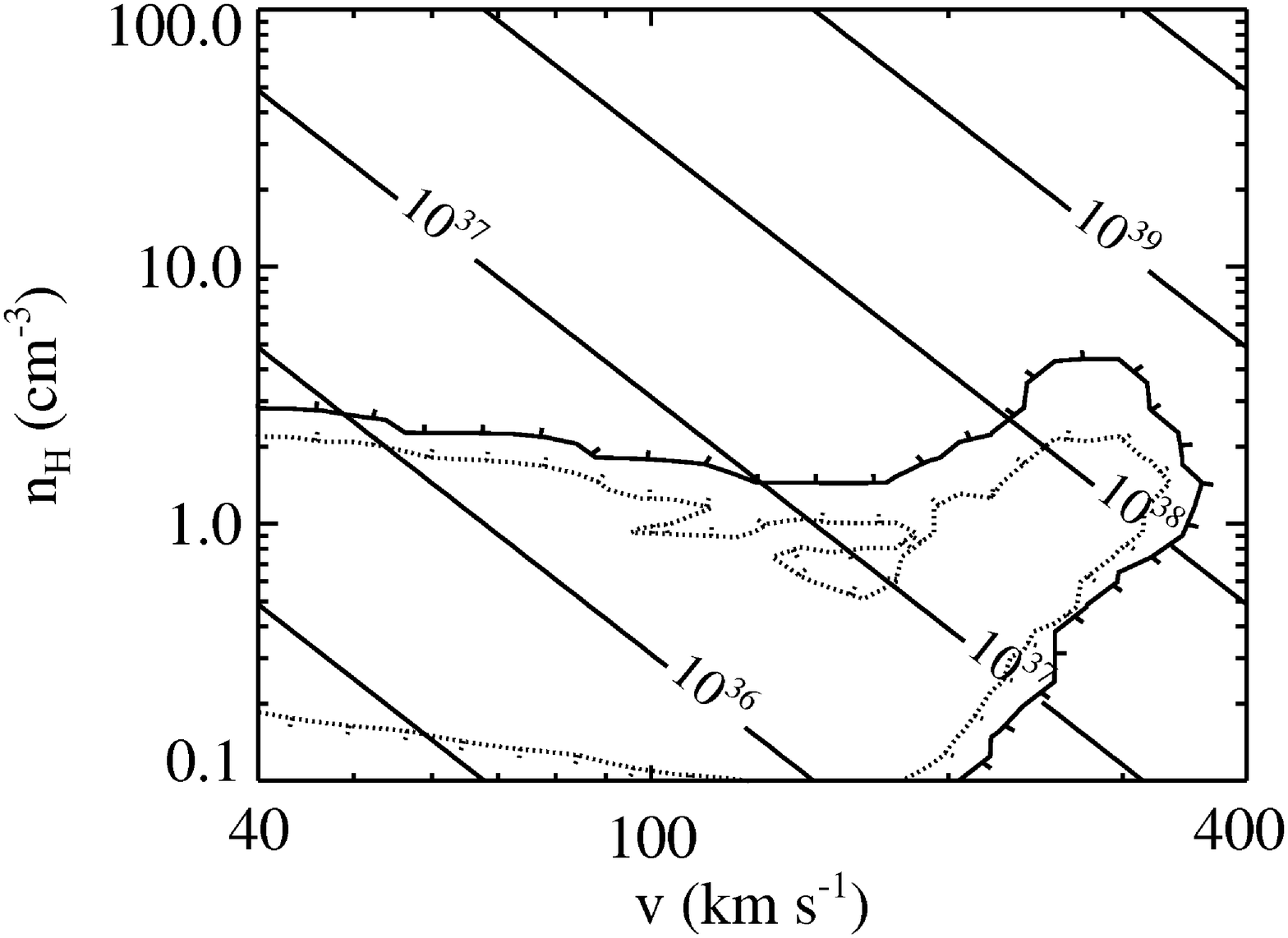}
\end{minipage}
&
\begin{minipage}{0.31\textwidth}
 	\includegraphics[trim = 0.8in 0.0in 0.4in 0.0in, clip, width=\textwidth]{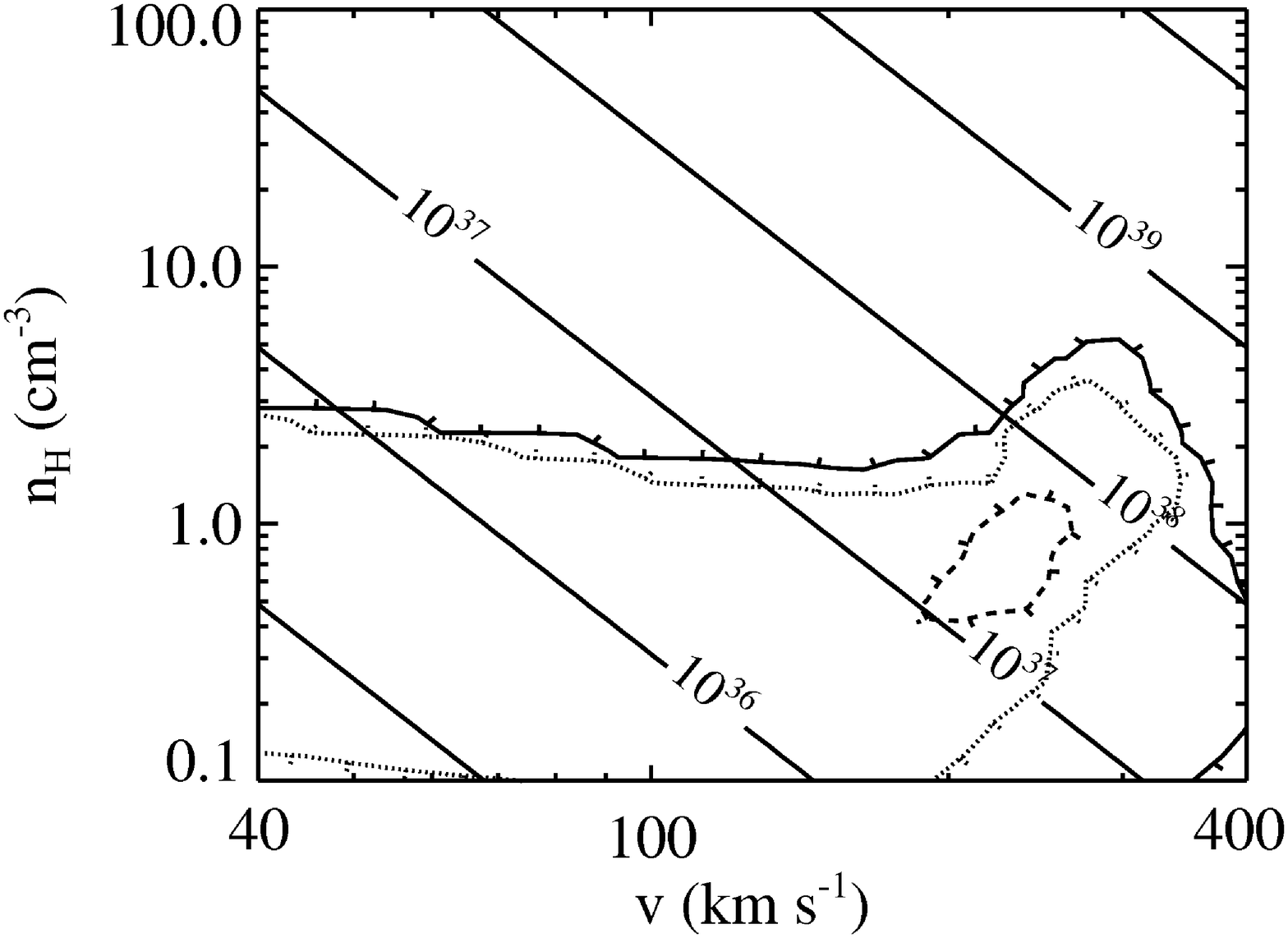}
\end{minipage}
\end{tabular}
 \caption{These contour plots summarize our analysis of the shock, which compares all of our multi-wavelength observational constraints to the MAPPINGS~III shock models.  The plots contain confidence contours calculated in three ways:  individual $1\sigma$ contours (left) and joint probability contours, where we do (right) and do not (middle) take into account systematic model uncertainties (see text for details).  In the left plot, a different linestyle represents each observational constraint:  [S~II] (6716~\AA~/~6731~\AA) (solid), [N~II] (6549~\AA~$+$~6583~\AA)~/~H$\alpha$ (dotted), [S~II] (6716~\AA~$+$~6731~\AA)~/~H$\alpha$ (short-dash), X-ray (dash-dot), radio (long-dash), and extinction-corrected H$\alpha$ surface brightness (dash-three dots).  The [O~III] (6549~\AA~$+$~6583~\AA)~/~H$\beta$ observations are too uncertain to appear on this plot.  Mutually excluded regions are shaded grey.  In the middle and right plots, we plot the joint $1\sigma$ (short-dash), $2\sigma$ (dotted), and $3\sigma$ (solid) contours.  Assuming that this is a jet-driven outflow, the diagonal contours in each plot correspond to the two-sided jet power, $P_{tot} \sin{(i)} / a$, needed to drive the shock in erg~s$^{-1}$ from Eq.~\ref{eq:pow}.  Downhill ticks have been provided in each plot to indicate the direction of the contour levels.}
 \label{fig:analysis}
\end{figure*}

To investigate the parameter landscape visually, we first
plotted the probability contours for each observable separately.  The
resulting plot is shown in the left panel of Fig.~\ref{fig:analysis}.
In order to minimize the effects of extinction, we followed standard
practices by considering ratios of lines in similar wavelength ranges,
namely: [S~II]/H$\alpha$\footnote{The [S~II] / H$\alpha$ ratio is a
particularly relevant discriminant for shock-heated gas.},
[N~II]/H$\alpha$, and [O~III]/H$\beta$.  The only diagnostic
significantly affected by extinction is the H$\alpha$ surface brightness;
we discuss the issue of a possible extinction correction error together
with the radio diagnostic below.

The figure shows 68.27 per cent confidence contours for each diagnostic and
indicates that a region of parameter space exists in which all
diagnostics are consistent with the data within $1\sigma$ confidence,
spanning the velocity range $100\,{\rm km\,s^{-1}}\la v_{\rm
  shock} \la 300\,{\rm km\,s^{-1}}$ and the density range
$0.2\,{\rm cm^{-3}} \la n_{\rm ISM} \la 2\,{\rm cm^{-3}}$.

A number of important features stand out in the plot:
\begin{itemize}
\item{The X-ray contours constrain the velocity to be below about
    $v_{\rm shock} \la 300\,{\rm km\,s^{-1}}$.  Higher velocities
    would lead to gas temperatures that would generate detectable
    X-ray emission.  Because the X-ray limit is physically very
    robust, it is essentially independent of the other model
    parameters.}
\item{All constraints except for the X-ray limits exclude low
    density/high velocity shocks.  The diagonal contour from roughly
    $v_{\rm shock} \approx 200\,{\rm km\,s^{-2}}$ and $n_{\rm ISM}
    =0.1\,{\rm cm^{-3}}$ to $v_{\rm shock} = 400\,{\rm km\,s^{-1}}$
    and $n_{\rm ISM} \approx 3\,{\rm cm^{-3}}$ demarcates the region of
    parameter space where the shock becomes radiative (to the left).
    To the right of this line, the shock never cools to temperatures
    where ionization species observed in the spectrum
    ([S~II],[N~II],[O~III]) exist at sufficiently high
    fractional abundance to match the observations.}
\item{The H$\alpha$ and radio contours are close to parallel, showing
the degeneracy in both diagnostics; whether or not we include either one
or both of these diagnostics does not bias our results.  This suggests that
our measurements are not significantly affected by multiple degenerate
possible systematic uncertainties (none of which we can better address
with the current data).  These systematic uncertainties include the
extinction correction for H$\alpha$, the lack
of non-thermal emission mechanisms (e.g. synchrotron) in the MAPPINGS III
models, and a lack of a limb-brightening correction for the radio (H$\alpha$
is not affected by limb-brightening because any optical fibers near the edge
of the shell are excluded from our analysis).}
\end{itemize}

In order to derive formal confidence intervals for the two parameters
of interest, we derived the joint probability distribution for all 7
observables.  We used the following approach:
\begin{enumerate}
\item{We derived the two-sided cumulative likelihood function
    $\mathcal{P}_{i}=\mathcal{P}(f_{i})$ for each parameter $f$, such
    that
    \begin{equation}
      \mathcal{P}(f) \equiv 1 - 2\left|{\int_{f_{\rm min}}^{f}df^\prime P(f^\prime)} -
          1/2\right|
    \end{equation}
    from the Monte-Carlo normalized probability distribution $P(f)$
    for parameter $f$ derived in Section \S\ref{section:optical_analysis}.  Note that
    $\mathcal{P}(f)$ peaks at 1 for the best fit value and approaches
    zero for large deviations from the best fit.  For the X-ray upper
    limit, we used the one-sided cumulative likelihood function
    $\mathcal{P}(f)=1 - \int_{0}^{f}df^\prime P(f^\prime)$.}
\item{We multiplied the two-sided cumulative likelihood for
    all seven parameters to derive a joint cumulative likelihood
    $\mathcal{P} \equiv \Pi_{i=1,7} \mathcal{P}_{i}$. This number is
    necessarily much smaller than one even for statistically
    satisfactory fits.}
\item{To evaluate the statistical significance of a given joint
    $\mathcal{P}$, we derive the distribution of $\mathcal{P}$ values
    expected for a set of seven independent observables (one of which
    is an upper limit), using Monte-Carlo simulations.  From the
    simulations, we derived the probability distribution to find, in a
    set of observations, a given $\mathcal{P}$ value.  We derived the
    expected $\mathcal{P}$ values for 68.27 per cent, 95.45 per cent, and 99.73 per cent
    confidence interval, which correspond to
    $\mathcal{P}=3.38 \times 10^{-4}, 5.95 \times 10^{-6}, 6.04 \times 10^{-8}$, respectively.
    The confidence intervals corresponding to these values are plotted
    in the contour plot of $\mathcal{P}$ shown in middle of Fig.~\ref{fig:analysis}}
\end{enumerate}

The middle panel of Fig.~\ref{fig:analysis} shows that even the best fit in our model grid
deviates from the observed data set by the equivalent of more than $1\sigma$.  However, to rule out the null hypothesis that the shell around
Cygnus~X-1 can be described by a shock model, we would traditionally
require a deviation of more than $3\sigma$ significance.  Thus, we
conclude that the observed shell is consistent with a plane parallel
shock model at roughly the $1.5\sigma$ level.

This lack of a statistically satisfactory fit is not surprising:  the
model we employ, while significantly more sophisticated than previous
approaches, necessarily makes some simplifying assumptions, which
introduce a considerable amount of systematic uncertainty.  For
example, given the ellipsoidal shape of the shell, the plane
parallel approximation breaks down for the inner shells of the shock.
Also, while the fibers used in this analysis are located towards the
projected center of the shell, they will still sample a range in shock
thicknesses due to inclination effects, as well as possibly a range in
external densities if the ISM near Cygnus X-1 is non-uniform.

In order to account for these and other sources of systematic
uncertainty, we re-evaluated the likelihood contours by increasing the
uncertainties in our parameter estimates.  We did this by assuming
that the best fit shock model can provide a statistically satisfactory
fit ($\chi^2 / \nu = 1$) if systematic
uncertainties are accounted for.  We derived the likelihood scaling
factor $\phi$ by assuming that a statistically satisfactory fit will
have a $\mathcal{P}_{50}$ value better than 0.5, compared to all
randomly drawn sets of data from the Monte Carlo distribution.  This
sets $\phi = \mathcal{P}_{50}/\mathcal{P}_{\rm max} = 6.24$.

We multiply the distribution of $\mathcal{P}$ values in our model grid
by $\phi$ to find the new confidence contours.  These are shown in the right
plot of Fig.~\ref{fig:analysis}.  By construction, we can now define
$1\sigma$, $2\sigma$, and $3\sigma$ confidence intervals.  While the $1\sigma$
contours constrain parameters very tightly, it is clear from the figure, the $2\sigma$ and
$3\sigma$ confidence contours allow a large region of parameter space.
From the figure, we can constrain the density to be $n_{\rm ISM}
<5\,{\rm cm^{-3}}$ and the velocity to be $v_{\rm shock} < 400\,{\rm km\,s^{-1}}$
with $3\sigma$ confidence.

We have an additional constraint on the velocity.  Since we do not resolve
the motion of the shock in our spectra, we place an upper limit on the radial
shock velocity of 250~km~s$^{-1}$.  Previous work indicates that the
inclination of the accretion disk is fairly low, but is difficult to determine
and fairly uncertain \citep{wiktorowicz09}:
$23.7^\circ$ \citep[from X-ray reflection spectra;][]{fabian12},
$27.1^\circ$\citep[from dynamical modeling;][]{orosz11},
$<55^\circ$ \citep[from H$\alpha$ tomography;][]{sowers98}, and
$\sim26-67^\circ$ \citep[from variability and a complilation of other techniques;][]{gleissner04}.
While the disk and jet could be misaligned from a natal kick
\citep{martin10}, this is highly unlikely because Cygnus~X-1 has a low space
velocity \citep[$v\approx21$~km~s$^{-1}$;][]{reid11}.  Given this information
regarding the inclination of the jet, our upper limits
for the shock velocity from the comparison to the MAPPINGS~III shock models
are consistent with this constraint.
This constraint is not robust because we may also not see the velocity split
in the lines simply because the emitting material at the shock front is too
faint in comparison to much more slowly moving post-shock gas.

Finally, we can convert the density and velocity limits to constraints
on the jet power.  We follow the approach taken when estimating the
power required to drive expanding cavities in galaxy clusters by AGN
jets \citep[e.g.][]{mcnamara07}.  Namely, we assume that the pressure inside the
shell is uniform and that the shell is roughly spherical.  In this
case, the expansion of the shell is well described by self-similar
models of wind driven bubbles \citep{castor75}, with a correction
proportional to $a/\sin(i)$, where $a$ is the aspect ratio of the
prolate ellipsoid and $i$ is the inclination.  Consistent with the
parameter range sampled, we assume the strong shock limit.  We also assume
the shock is isothermal, while the actual shock is radiative, which can
cause a discrepancy between the actual and analytical estimate of the
thickness of the shock used in the following equation.  For a
given set of $n_{\rm ISM}$ and $v_{\rm shock}$, the power required to
drive the observed shell can be estimated
\begin{equation}
\label{eq:pow}
  P_{\rm tot} = \frac{a}{\sin{(i)}}\frac{11}{27}8\pi
  R_{\rm s}^2 \rho_{\rm ISM} v_{\rm s}^3
\end{equation}
where we assumed a monatomic relativistic ideal gas equation of state
($\gamma = 4/3$) for the interior and a non-relativistic monatomic
ideal gas equation of state for the ISM ($\gamma = 5/3$).

From the joint probability distribution, we find the formal best-fit density,
velocity, and two-sided jet power:  $0.74_{-0.32}^{+0.57}$~cm$^{-3}$,
$220_{-30}^{+50}$~km~s$^{-1}$, and $2.6_{-1.7}^{+4.2} \times 10^{37}$~erg~s$^{-1}$,
respectively.  Contours of constant $P_{\rm tot}\sin{(i)}/a$, where $i=90^\circ$
and $a=1.3$ (measured from the H$\alpha$ image) are plotted on all plots in
Fig.~\ref{fig:analysis}.  Our results of this analysis are summarized in Table~\ref{table:results}.

\begin{table}
  \centering
  \caption{Summary of Results}
  \begin{threeparttable}
    \begin{tabular}{lll} \hline \hline
      Parameter                              & Individual$^a$ & Joint$^b$    \\ \hline
      Velocity (km~s$^{-1}$)                 & 80~--~300      & 190~--~270   \\
      Pre-shock density (cm$^{-3}$)          & 0.16~--~1.80   & 0.42~--~1.31 \\
      Jet power$^c$ ($10^{37}$~erg~s$^{-1}$) & 0.04~--~10     & 0.9~--~6.8   \\ \hline
    \end{tabular}
    \begin{tablenotes}
   		\item[a] One sigma confidence intervals corresponding to the left plot of Fig.~\ref{fig:analysis}.
   		\item[b] One sigma confidence intervals corresponding to the right plot of Fig.~\ref{fig:analysis}.
   		\item[c] Two-sided.
    \end{tablenotes}
  \end{threeparttable}
  \label{table:results}
\end{table}

\section{Discussion}
\label{section:discussion}

In light of our new observations and analysis, we revisit the interpretation of the Cygnus~X-1 outflow.  We have put tighter and more physically-motivated limits on the pre-shock velocity (the shock velocity measured in the lab frame; $v$), pre-shock density ($n$), and power needed to drive the shockwave ($P_{\rm tot}$; Table~\ref{table:results}) compared to previous work \citep{gallo05,russell07}.  Using these new constraints, we explore the source of power for the shockwave in this section.

\subsection{The Black Hole Jet}

First, the near perfect alignment between the small-scale radio jet and large-scale shockwave, which are both at a position angle $21^\circ$\footnote{Given uncertainties in the position angle of a few degrees for each, a chance orientation is only a few per cent.}, leads us to first investigate the case where the shockwave is driven exclusively by the BH jet.  For comparison to models and other (XRB and AGN) observations, we calculate the efficiency at which accreted power is converted to other forms of power.  

The bolometric luminosity is commonly used to trace the accretion rate:
\begin{equation}
L_{bol} = \eta \dot{m} c^2
\end{equation}
where $\eta$ is the conversion efficiency \citep[e.g.][]{novikov73}.  However, the total power in the accretion process is not only radiated; some (potentially a large fraction) is advected or is ejected in a jet or disk wind (the kinetic power).  Therefore, we more explicity specify how the accretion power, $P_{acc}$, can be transformed into other forms of power \citep[following][]{jester05}:
\begin{equation}
\begin{split}
&P_{rad} = \epsilon \delta \dot{m} c^2 = L_{bol} \\
&P_{adv} = \alpha \delta \dot{m} c^2 \\
&P_{kin} = \kappa \delta \dot{m} c^2 \\
\end{split}
\end{equation}
where
\begin{equation}
\begin{split}
&P_{acc} \equiv \delta \dot{m} c^2 \\
&\eta \equiv \epsilon \delta \\
&\epsilon + \alpha + \kappa \equiv 1 \\
\end{split}
\end{equation}
and where the latter relation is for a non-spinning BH.  In this representation, $\delta$ represents the fraction of rest-mass energy liberated by the accretion process, which scales with the radius of the innermost stable circular orbit and is fundamental to the spacetime metric.  For a non-spinning (Schwarzschild) BH, $\delta = 0.06$ but it can be as high as $\delta = 0.42$ for a rapidly spinning (Kerr) BH.  The relative values of $\epsilon$, $\alpha$, $\kappa$ depend on the accretion flow solution.

In the case of Cygnus~X-1, we estimate $L_{bol}$ using the 0.1~--~200~keV hard state luminosity, $L_{bol} = 3 \times 10^{37}$~erg~s$^{-1}$ \citep[e.g.][]{disalvo01}, which is $0.01 L_{Edd}$ for a 14.8~M$_\odot$ BH.  We adopt the above value for $L_{bol}$ with a couple of caveats.  First, we do not include the recently observed MeV spectral tail in $L_{bol}$, which would only be a small correction \citep[but is important for understanding jet physics;][]{zdziarski12}.  Second, we have made no duty cycle correction to the X-ray bolometric luminosity because only a very small correction is implied (much smaller than the uncertainties in the other relevant parameters):  Cygnus X-1 has spent most of the time \citep[$\sim 80$ per cent;][]{grinberg13} in approximately the last two decades in this state and this correction is likely much smaller than long-term (over thousands of years) changes in the accretion rate.

Next, we can use these constraints with the parameterization above to explore the kinetic efficiency, which is the jet efficiency if the jet is driving the outflow (i.e. $\kappa \rightarrow \kappa_{jet}$).  Cygnus~X-1 completes a small track in the hardness-intensity diagram in intermediate states, indicating that it never strays too far from the transition between its intermediate soft and hard states \citep[$\dot{m}_{crit}$; e.g.][]{torii11} like low-mass XRBs typically do \citep[e.g.][]{miyamoto95,smith02,maccarone03a}.  In this regime, if the accretion flow is characterized by an advection-dominated accretion flow, the advection efficiency is expected to be small \citep[$\alpha \sim 0.35$;][]{esin97}.  This is the efficiency we assume below.\footnote{The advection efficiency is highly uncertain because the advection-dominated accretion flow model is not a full description of the accretion flow and because it depends on the radiative efficiency, which is degenerate with the accretion rate.  It cannot be too large because the accretion rate would then have to be unphysically large.  However, placing a stringent upper limit on the accretion rate is very difficult because the fraction of the wind that is focused into the accretion stream is highly uncertain.}  In its soft state, the radio emission drops dramatically \citep[e.g.][]{tananbaum72}, indicating that the jet has been mostly quenched \citep{fender06}.  At the same time, the X-ray luminosity increases by factors of a few, making $\epsilon \sim 1$.  In its hard state, $P_{rad}$ and $P_{kin}$ are approximately equivalent, meaning that $\epsilon \sim \kappa \sim \alpha$.  This corresponds to an average accretion rate of $2 \times 10^{-8}$~M$_\odot$~yr$^{-1}$.  In this most likely case, there is enough kinetic power in the jet to power the outflow.

However, we cannot rule out a large range of outflow powers with high significance.  In particular, considering up to the $3\sigma$ limit gives $\kappa / \epsilon < 7$.  Such a large jump in the accretion power needed to fuel the jet at the state transition could not be hidden in bolometric corrections \citep{maccarone05b}, and increasing the advection efficiency only exacerbates the problem.  This would imply that there is not enough accretion energy to power the outflow via the jet unless $\epsilon + \alpha + \kappa > 1$.  In this case, the jet would have to extract considerable amounts of rotational energy from the BH \citep{blandford77}.  This would support the claim that Cygnus~X-1 is rapidly spinning \citep{fabian12,gou14}.  This would also provide support for the claim that the jet power is higher for more rapidly rotating BHs based on an observed correlation between the peak radio fluxes of XRBs and the spins fitted from disk continuum modeling \citep{narayan12}.  However, other studies of ballistic jet ejections and radio/X-ray correlations from low/hard state BHs do not find strong effects of BH spin \citep{fender10,russell13}.  These contradictory studies together with our measurements lead us to conclude that we have no reason to favor the idea that the high efficiency of jet production is due to different values of the BH spin.

The BH jet can also be placed in the context of the standard model for BH jets, which may not be complete.  \cite{heinz06} carried out a detailed comparison of the observed jet power to that estimated from the canonical model for flat radio emission from jet cores \citep{blandford79}.  \citeauthor{heinz06} found a dramatic difference between the small- and large-scale jet power estimates (observed, large-scale jet power is too large) and could only reconcile them through some combination of non-standard factors:  a small filling factor, a mass-loaded jet \citep[e.g.][]{sikora00,diaztrigo13}, a contribution of other sources of power to the large-scale shockwave (see sections \ref{section:supernova} and \ref{section:ostar}), and/or a some fundamental issue with the jet model itself.

Constructing even a small, comparative sample of sources to place this analysis in context is challenging.  There are only a couple of other sources in the Milky Way with well-measured, large-scale shockwaves:  Circinus~X-1 and SS~433, but the large-scale features have a considerably different morphology from that of Cygnus~X-1.   Instead of Cygnus~X-1's one-sided, parsec-scale bow shock, these two systems are embedded in their natal supernova remnants, making them unusually young XRBs (a few $\times 10^3$~yr for Circinus~X-1 --- \citealt{heinz13} and $\sim 2 \times 10^4$~yr for SS~433 --- e.g. \citealt{goodall11}).  These binaries have dual, parsec-scale termination shocks of the bipolar jets referred to as ``ears" for SS~433 \citep[e.g.][]{geldzahler80,elston87,dubner98} or ``caps" \citep{heinz07b,sell10}.  The implied kinetic power of the jets in both systems is enormous:  $\sim 10^{39}$~erg~s$^{-1}$ for SS~433 \citep{panferov97} and $3 \times 10^{35}$~--~$2 \times 10^{37}$~erg~s$^{-1}$ for Circinus~X-1 \citep{sell10} but could be as high as $\sim 10^{39}$~erg~s$^{-1}$ \citep[based on what could be a protrusion of the jet on the south side of the supernova remnant;][]{heinz13}.

These two systems also have considerably different accretion flows compared to Cygnus~X-1's well-characterized BH binary wind-fed system, which has spent most of the time in the last few decades in the hard state where a steady jet is seen \citep{stirling01} and expected to be found \citep{fender04}.  SS~433 appears to have a supercritical, heavily extincted accretion flow with considerable UV emission and strong winds surrounding what is a likely but not a securely determined $\sim 10$~M$_\odot$ BH \citep{fabrika04}.  This system may more closely resemble the population of ultra-luminous X-ray sources seen in nearby galaxies \citep{begelman06,fabrika06}.  On the other hand, Circinus X-1 is a confirmed neutron star \citep{linares10} XRB in a highly eccentric orbit with very strong variability, showing both Z- and Atoll-like behavior \citep[e.g.][]{shirey99}.  In its very low state, the system does exhibit a steady jet \citep{calvelo12a,calvelo12b}.  The fact that it is not easily classified with the rest of the neutron star binary population can now be attributed to the fact that it is in a unique evolutionary phase as the youngest neutron star XRB in the galaxy \citep{heinz13}.  These qualitative differences can bring about considerable quantitative systematic uncertainties in $\epsilon$ and $\alpha$, and, therefore, $\kappa_{jet}$ (these could change by an order of magnitude or more between states because the variability is so strong).  Because $\kappa_{jet}$ is so poorly constrained, we choose not to try to estimate it to compare to Cygnus X-1.  We discuss some of these issues in the context of Cygnus X-1 in section \ref{section:future}.

Finally, the Cygnus X-1 jet may also be better understood when comparing our results to samples of AGN where the total jet power can be measured \citep[e.g.][]{allen06}.  Such comparisons could be problematic because $L_X$ does not provide a good estimate for $P_{acc}$ in supermassive BHs as it does for XRBs \citep{pellegrini05}.  Nevertheless, \cite{merloni07} and \cite{kording08} both found the estimates of the scaled jet power compared to the accretion power for Cygnus~X-1 to be consistent with those of their AGN samples, albeit with large uncertainties comparable to the uncertainties in the jet power that we find in this work.  In contrast, \cite{king13} claimed to find a discrepancy between the jet and wind-scaling correlations with the bolometric luminosity and the values found for Cygnus X-1 using different scaling relations.

In the next two subsections, we consider other possible sources for the observed shockwave.

\subsection{A Supernova Remnant?}
\label{section:supernova}

We consider whether the data suggest that this shockwave could be the remnant of the supernova that created the Cygnus~X-1 BH.  In this case, the distance from Cygnus X-1 to the shockwave and the limits on the velocity imply that the shockwave would likely be near the transition of the Sedov-Taylor \citep{taylor50,sedov59} and ``snow-plough" phases.  Given our constraints on the density ($n < 5$~cm$^{-3}$), size ($R_{max,avg} < 6$~--~9~pc), and velocity ($v < 400$~km~s$^{-1}$) of the shockwave (we do not expect that our limits on these three parameters are strongly model-dependent), the upper limit on the supernova energy in these phases is $\la 1 \times 10^{50}$~erg \citep[e.g.][]{vink12}.  Because $E$ is sensitive to $R$ ($E \propto R^3$) and the shockwave is asymmetric and possibly inclined, the uncertainty in this upper limit is approximately a factor of 2~--~3.  Nevertheless, this upper limit is approximately an order of magnitude fainter than the typical supernova energy of $\sim 10^{51}$~erg \citep[e.g.][]{chevalier77,korpi99}.

This leads us to consider a couple possibilities.  First, the shockwave could be the remnant of a low-energy explosion, where most of the mass was promptly swallowed by the BH, also explaining the low proper motion of the binary.  This has already been suggested by \cite{mirabel03}, but in a different context before the shockwave had been discovered.  \citeauthor{mirabel03} considered the kinematics of the entire region including the proper motion of Cygnus~X-1 relative to the nearby Cyg OB3 association.  They suggested that the BH progenitor, a $\ga 40$~M$_\odot$ star, exploded within the nearby H~II region a few million years ago, producing a low-energy shockwave in the H~II region or none at all.  Assuming, instead, that the current young (a few $\times 10^4$~yr) shockwave is the low-energy explosion that they suggest, could explain the unusual, but not implausible morphology \citep[e.g. N132D;][]{blair00} and the lack of unusual metal abundance patterns (e.g. oxygen, nitrogen, sulfur) expected for typical supernovae.  Since supernova blastwaves can sometimes have lower energies as constrained above \citep[e.g.][]{jones98}, this scenario remains an intriguing possibility.

Another simpler possibility is that this could be a supernova remnant unrelated to Cygnus~X-1, seen in chance projection at a larger distance.  The shockwave would be much more extended in this case, bringing the required energy more in line with a typical supernova energy.  As this is close to a nearby star-forming region, Cyg~OB3, perhaps another supernova occurred behind Cygnus~X-1 in the more distant past.  We cannot rule out this possibility.  However, given the excellent alignment with the BH jet, the unusual morphology, and the lack of unusual abundance patterns as stated above, the simplest, most likely explanation is that the shockwave is associated with Cygnus~X-1.

\subsection{The O-star}
\label{section:ostar}

The O-star companion in the binary can have a strong effect on the ISM and shockwave near Cygnus~X-1.  First, the photoionizing spectrum could have a large impact on the ISM ahead of and behind the shockwave.  Given that the Str\"{o}mgren sphere of the O-star is $\approx 40$~pc using the limits on the pre-shock density we measure, the O-star should photoionize the gas throughout this entire shocked region \citep[e.g.][]{tielens05}.  Our model comparisons to the data, which do take pre-photoionization into account, provide some insight here.  While neither the data nor the models are detailed enough to make precise statements concerning the level of ionizing flux at the shock front and they do not enable us to rule out any of the scenarios we consider, we do find that the shock models consistently prefer a relatively low fractional ionizing O-star flux.  The data are almost always best-fit with the $(1/16) F_{O-star}$ models (see Table~\ref{table:model_params}).  This is puzzling and could be explained by a number of possibilities:  there is considerable attenuation of ionizing radiation (e.g. by dust) in between the O-star and the shockwave, the shockwave is considerably inclined to the line of sight, the shockwave is unassociated with Cygnus~X-1, or limitations of the one dimensional MAPPINGS~III shock models are the cause of this discrepancy.

We also consider the role of the O-star wind.  We use the measured wind mass-loss rate \citep[$2.5 \times 10^{-6}$~M$_{\sun}$~yr$^{-1}$;][]{herrero95,gies03} and the measured characteristics of the companion star in Cygnus~X-1 \citep[$T_{eff} = 28.0 \pm 2.5$~kK, log~$g \ga 3.00 \pm 0.25$, and $L_* = 2.0 \times 10^5 L_{\sun}$,][]{caballero-nieves09} together with a model of theoretical mass loss rates of isolated O- and B-stars \citep{vink00}.  This model includes a wind efficiency/performance number, $\eta$, that describes the fraction of the momentum of the radiation that is transferred to the ions in the wind.  For these input values, $\eta \sim 0.1$~--~0.2, implying a spherical wind power of $\sim 10^{38}$~erg~s$^{-1}$.  Unless the wind is somehow focused, $\sim15$ per cent of the wind power is directed toward the shockwave.  This is consistent with the measured power needed to drive the shockwave, suggesting that the O-star could be an important driving force behind the shockwave.

To investigate the possibility of O-star winds driving the shock, we consider a well-known, extreme example of a system with O-star wind-blown shocks, Eta Carina.  In an outburst over 100 years ago, which could have been confused for a supernova explosion \citep{davidson12a}, this binary double O-star has driven powerful bipolar outflows \citep{davidson12b}.  Multiple observations have revealed high outflow velocities of thousands of kilometers per second, well beyond the escape velocity of the system \citep{behar07,ishibashi99,smith08}.  Various wind models over the past couple of decades have attempted to reproduce the powerful, bipolar outflows of Eta Carina \citep[e.g.][]{hillier97,gonzalez04,smith13} with varying degrees of success.  However, no model yet exists to take into account the unique effects of a BH companion instead, as in the case of Cygnus~X-1.

\subsection{Summary and Future Prospects for Determining the Source of the Outflow}
\label{section:future}

While we have concluded that a supernova origin for the outflow is unlikely, we have found that the O-star wind and BH jet both remain strong possibilities and likely work together to create the outflow and influence the state of the gas on both sides of the shock front.  For instance, recent simulations clearly show how the O-star wind can be strongly affected by the dynamics of the gas very near the BH \citep{hadrava12,yoon14}.  Additional thermal heating, photoionization, bulk gas motions, clumping, and additonal turbulence likely present in a BH binary system may work to alter outflows.  In addition, the winds of the previous and current O-stars in the binary can produce a variable density profile in the nearby ISM, whereas the MAPPINGS~III shock models assume a uniform density ISM.  Detailed three dimensional models are needed that take into account these effects and translate how they affect large-scale outflows.

In addition, a better, more robust determination of the large-scale outflow power, including a more robust lower limit on the outflow velocity, would help to constrain the outflow.  However, this is a challenging prospect because of considerable observational and theoretical hurdles that add large uncertainties:
\begin{itemize}
\item If the jet is driving the outflow, it could be injecting most of its kinetic power in different modes (steady versus discrete jet ejections).  For instance, defining $\dot{m}$ is problematic if most of the accreted mass is lost in a wind instead of a jet or if most of the injected energy in the jets is during a short-lived powerful outburst where the $\epsilon$, $\alpha$, or $\kappa$ could vary radically.  There is not much evidence for this in the last few decades while Cygnus~X-1 has been well-monitored, but this is only a small fraction of the lifetime of the BH.  The upper limit on the velocity ($v<400$~km~s$^{-1}$) implies that the outflow was launched $\ga 10000$~yr ago.
\item How energy is injected into large-scale lobes and shockwaves may not be fully understood.  We may be considerably underestimating ($\sim 6\times$) how much energy is needed to create the large-scale lobe north of Cygnus~X-1, which is analyzed similarly to radio lobes in galaxy clusters \citep{binney07}.
\item Despite the complexity of the 1-D shock models we employ that have been developed over many decades, the implementation of future 3-D simulations with more physical processes that are tracked and understood from the launching point to the large-scale outflow could considerably affect our interpretation and measurements.  This includes developing a better understanding how winds and jets might couple together to drive large-scale outflows.  Such simulations are challenging to develop and much more computationally expensive to run than the tens of thousands of computer hours for the 1-D shock models we used.
\item Much of the uncertainty in the outflow power arises from the our inability to constrain many of the shock model parameters tracing properties of the shock and ISM, most of which we marginalized over (see Table~\ref{table:model_params}):  inclination (and distance from Cygnus~X-1), density, velocity, magnetic field strength, abundance, and O-star ionizing flux.  Unfortunately, placing improved observational constraints on these parameters is very difficult and expensive in terms of the observing time required.
\end{itemize}

One solution could be to search for similar shockwaves around other XRBs, especially in cases where the stellar companion would not have a strong wind.  If other systems with large-scale shockwaves are found, this methodology could be applied again to determine the outflow power to within a factor of a few.  However, this could be problematic for a couple reasons.  First, if the shockwave north of Cygnus~X-1 is mostly powered by the O-star wind, then such shockwaves may not exist in most cases.  Second, such shockwaves would exist at a range of random ages.  As noted in Section \ref{section:models}, our data are heavily constrained by models to the left of the cooling line where the temperature is low enough that low ionization species are observable.  However, a large fraction of the optical radiative output of the shockwaves are in emission lines.  This implies that such shockwaves may not be observable via optical lines for most of their lifetimes because the cooling time is so short.  However, the shock could be re-energized with the continuous injection of energy by the BH jet, extending the cooling time.  In any case, we conclude that making substantial progress in firmly determining and better constraining the source of this and other similar outflows beyond initial, rough, simple estimates may require large strides in theoretical understanding and observational capabilities.

\section{Conclusions}
\label{section:conclusions}

We have presented an in-depth analysis of the parsec-scale outflow of Cygnus X-1 by comparing multiwavelength observations to detailed shock models.  We place physically-motivated results on the shock velocity, the pre-shock density, and the jet power.  With these new results, we discuss the complicated nature of the outflow and the list some possibilities for the source of the outflow power.  We find that, unless the shockwave was formed in an unusual supernova explosion, it is unlikely to be a supernova remnant.  The morphological alignment to the relativistic radio jet suggests that it plays an important or central role in driving the shockwave.  However, the O-star could also play an important role in creating a wind-blown bubble producing the large-scale shock.  We conclude that a significantly better characterization of the outflow appears to be a daunting task because it requires much better constraints on multiple observational and theoretical systematic uncertainties.

\section*{Acknowledgments}

We thank Brent Groves and Mike Dopita for helping us with setting up and running the MAPPINGS~III shock code.  SH and PS acknowledge support through NASA/Chandra grant G07-8040X and NSF grant AST-0908690.

\bibliographystyle{mn2e}

\bsp

\label{lastpage}

\end{document}